\documentclass[
reprint, 
superscriptaddress,
amsmath,
amssymb,
aps,
prl,
longbibliography,
]{revtex4-2}

\usepackage{graphicx}
\usepackage{dcolumn}
\usepackage{physics}
\usepackage{gensymb}
\usepackage{bm}
\usepackage{xcolor}
\usepackage{footmisc}
\usepackage{siunitx}
\usepackage{tabularx}
\usepackage{comment}
\usepackage{hyperref}

\usepackage[normalem]{ulem}

\begin{document}

\preprint{APS/123-QED}

\title{Temperature-dependent spin-lattice relaxation of the nitrogen-vacancy spin triplet in diamond}

\author{M.~C.~Cambria}
 
\affiliation{Department of Physics, University of Wisconsin, Madison, Wisconsin 53706, USA}

\author{A.~Norambuena}

\affiliation{Centro de Optica e Informaci\'on Cu\'antica, Universidad Mayor, Camino La Pir\'amide 5750, Huechuraba, Santiago, Chile}

\author{H.~T.~Dinani}
 
\affiliation{Centro de Investigaci\'on DAiTA Lab, Facultad de Estudios Interdisciplinarios, Universidad Mayor, Santiago, Chile}

\affiliation{Escuela Data Science, Facultad de Ciencias, Ingenier\'{i}a  y Tecnolog\'{i}a, Universidad Mayor, Santiago, Chile}

\author{G.~Thiering}

\affiliation{Wigner Research Centre for Physics, P.O. Box 49, 1525 Budapest, Hungary}

\author{A.~Gardill}

\author{I.~Kemeny}

\author{Y.~Li}
 
\affiliation{Department of Physics, University of Wisconsin, Madison, Wisconsin 53706, USA}

\author{V.~Lordi}
 
\affiliation{Lawrence Livermore National Laboratory, Livermore, CA, 94551, USA}

\author{\'A.~Gali}

\affiliation{Wigner Research Centre for Physics, P.O. Box 49, 1525 Budapest, Hungary}

\affiliation{Department of Atomic Physics, Institute of Physics, Budapest University of Technology and Economics, M\H{u}egyetem rakpart 3., 1111 Budapest, Hungary}

\author{J.~R.~Maze}
 
\affiliation{Instituto de F\'isica, Pontificia Universidad Cat\'olica de Chile, Casilla 306, Santiago, Chile}

\affiliation{Centro de Investigaci\'on en Nanotecnolog\'ia y Materiales Avanzados, Pontificia Universidad Cat\'olica de Chile, Santiago, Chile}

\author{S.~Kolkowitz}
\email{kolkowitz@wisc.edu}
 
\affiliation{Department of Physics, University of Wisconsin, Madison, Wisconsin 53706, USA}

\date{\today}
\begin{abstract}
Spin-lattice relaxation within the nitrogen-vacancy (NV) center's electronic ground-state spin triplet limits its coherence times, and thereby impacts its performance in quantum applications. We report measurements of the relaxation rates on the NV center's \(\ket{m_{s}=0} \leftrightarrow \ket{m_{s}=\pm 1}\) and \(\ket{m_{s}=-1} \leftrightarrow \ket{m_{s}=+1}\) transitions as a function of temperature from 9 to 474 K in high-purity samples. We show that the temperature dependencies of the rates are reproduced by an \textit{ab initio} theory of Raman scattering due to second-order spin-phonon interactions, and we discuss the applicability of the theory to other spin systems. Using a novel analytical model based on these results, we suggest that the high-temperature behavior of NV spin-lattice relaxation is dominated by interactions with two groups of quasilocalized phonons centered at 68.2(17) and 167(12) meV.
\end{abstract}

\maketitle


\textit{Introduction.}---The nitrogen-vacancy (NV) center is a point defect in diamond that has become a promising platform for quantum technologies ranging from nanoscale magnetic resonance imaging \cite{rugar2015proton, abobeih2019atomic} and electric field sensing \cite{iwasaki2017direct, barson2021nanoscale} to quantum information processing \cite{bradley2019ten, pompili2021realization, abobeih2022fault}. The NV center owes much of its appeal to its crystalline host, which allows the system to be placed near other materials or integrated into larger devices without the need for trapping or cooling \cite{vool2021imaging, patel2016efficient}. The NV center's solid-state environment also presents significant challenges, however. Interactions between the NV center's electronic spin and phonons in the surrounding crystal lattice drive spin-lattice relaxation \cite{norambuena2018spin}, also called spin-phonon relaxation \cite{xu2020spin, lunghi2022}, 
fundamentally limiting the system's achievable electronic spin coherence times and therefore its performance in quantum applications. Accordingly, several works have explored the temperature dependence of phonon-limited relaxation on the single-quantum transition between the \(\ket{m_{s}=0}\) and \(\ket{m_{s}=\pm 1}\) levels of the NV's ground-state electronic spin triplet \cite{redman1991spin, takahashi2008quenching, jarmola2012temperature, norambuena2018spin}. Although relaxation on the double-quantum transition between the \(\ket{m_{s}=-1}\) and \(\ket{m_{s}=+1}\) levels also limits coherence times, phonon-limited relaxation on this transition has been less thoroughly studied, leaving maximum achievable NV coherence times unknown for a wide temperature range.

To our knowledge, the only systematic measurements of phonon-limited double-quantum relaxation to date are Refs.~\cite{cambria2021state} and \cite{lin2021temperature}. In Ref.~\cite{cambria2021state}, we found that double-quantum relaxation occurs roughly twice as fast as single-quantum relaxation in the phonon-limited regime at room temperature. In Ref.~\cite{lin2021temperature}, Lin et al. present measurements of the single- and double-quantum relaxation rates between room temperature and 600 K, but several discrepancies with prior works suggest that these results cannot be easily generalized to describe NV spin-lattice relaxation in high-purity samples (Fig. S2 in \cite{SupplementalMaterial}). The temperature dependence of phonon-limited double-quantum relaxation has not been characterized below room temperature.

In this Letter, we present measurements of the temperature dependence of the single- and double-quantum relaxation rates of the nitrogen-vacancy center's electronic ground-state spin triplet in high-purity samples from 9 to 474 K. 
In contrast to prior theoretical understanding \cite{norambuena2018spin}, we argue that Raman scattering of phonons in the NV center arises mainly due to second-order, rather than first-order, spin-phonon interactions. We develop an \textit{ab initio} framework for calculating the relaxation rates that result from these interactions and we suggest similar spin systems that are subject to the same theoretical arguments. 
Informed by the presence of two vibrational resonances in the NV spin-phonon spectral function, we develop an analytical model of NV spin-lattice relaxation in which transitions are driven by two strongly coupled effective phonon modes at energies of 68.2(17) and 167(12) meV. Additionally, we use our experimental results to calculate the temperature-dependent limits to NV electron spin coherence times imposed by spin-lattice relaxation for superpositions in both the single- and double-quantum subspaces, and we compare these limits to the longest coherence times reported in literature. 

\begin{figure}[t]
\includegraphics[width=0.48\textwidth]{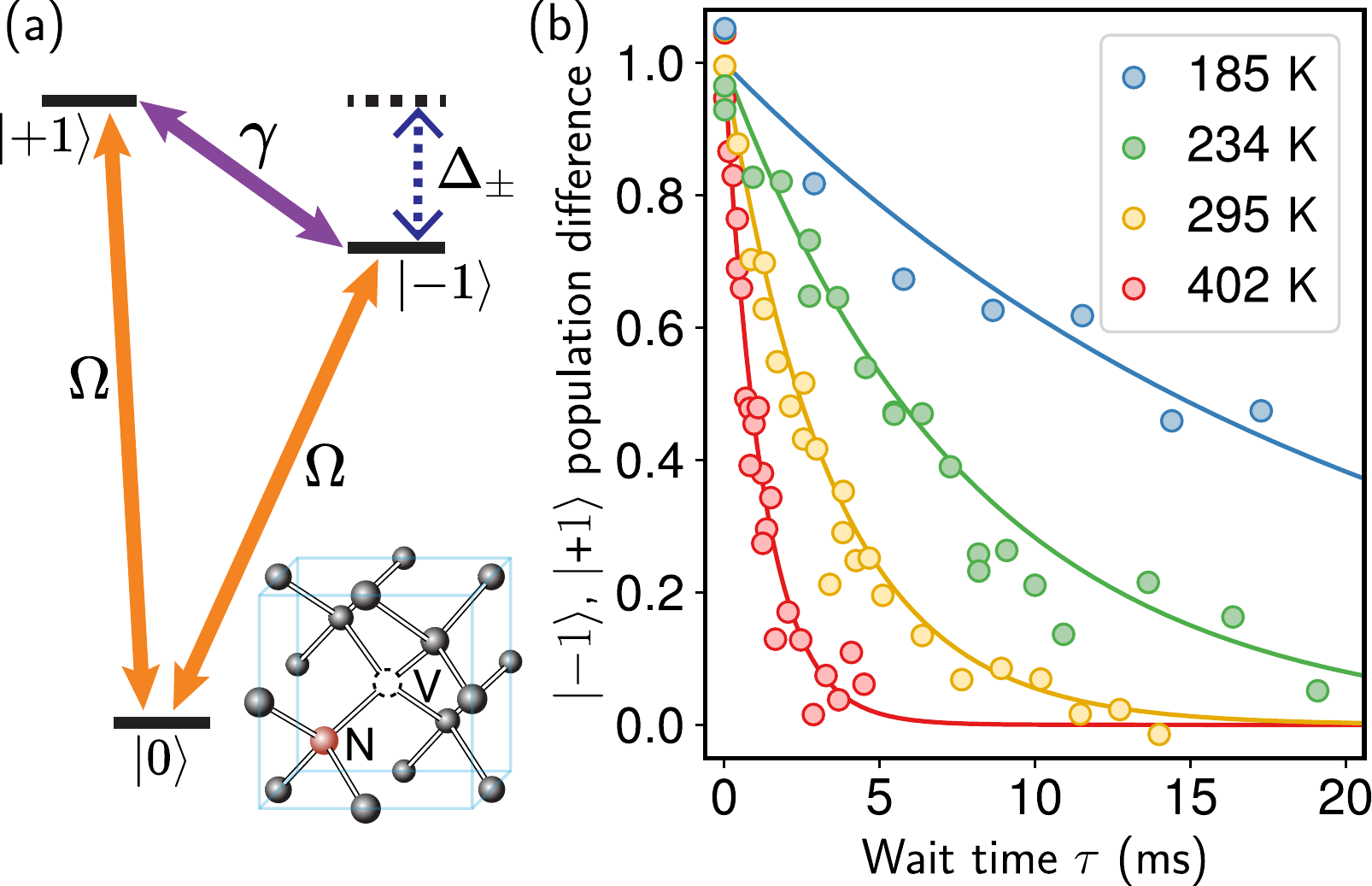}
\caption{\label{fig:introduction} 
Phonon-limited spin relaxation in nitrogen-vacancy centers.
(a) Level-structure of the NV\(^{-}\) ground-state electronic spin triplet. The zero field splitting raises \(\ket{\pm 1}\) above \(\ket{0}\) by 2.87 GHz at room temperature; a static magnetic field lifts the \(\ket{\pm 1}\) degeneracy. Relaxation between \(\ket{0}\) and \(\ket{\pm 1}\) (\(\ket{-1}\) and \(\ket{+1}\)) occurs at rate \(\Omega\) (\(\gamma\)). 
Inset: Nitrogen-vacancy defect within a carbon lattice.
(b) Effect of temperature on the curves used to extract the relaxation rate \(\gamma\). Data points show the difference between population in \(\ket{+1}\) and \(\ket{-1}\) after time \(\tau\) following initialization in \(\ket{+1}\) in sample A. Curves are single-exponential fits with decay rate \(2\gamma + \Omega\). 
}
\end{figure}

\textit{Experimental methods.}---Figure~\ref{fig:introduction}a shows the level structure of the NV's ground-state electronic spin triplet, where we abbreviate \(\ket{m_{s}=0}\) (\(\ket{m_{s}=\pm 1}\)) to \(\ket{0}\) (\(\ket{\pm 1}\)). Relaxation on the \(\ket{0} \leftrightarrow \ket{\pm 1}\) single-quantum transitions occurs at a common rate \(\Omega\); relaxation on the \(\ket{-1} \leftrightarrow \ket{+1}\) double-quantum transition occurs at rate \(\gamma\). Prior studies of phonon-limited spin relaxation in NVs have focused on the lifetime of \(\ket{0}\) under the assumption that \(\ket{0}\) and either \(\ket{-1}\) or \(\ket{+1}\) can be considered a qubit and the third level of the spin triplet can be neglected, tacitly assuming $\gamma=0$ \cite{redman1991spin, takahashi2008quenching, jarmola2012temperature, norambuena2018spin}. These prior works use \(T_{1}\) to denote the lifetime of \(\ket{0}\), which is related to the rates used in this work by \(T_1=1/(3\Omega)\).

The experimental methodology we employ to measure \(\gamma\) and \(\Omega\) is similar to that of prior works \cite{myers2017double, gardill2020fast, cambria2021state}. Optical polarization and state-selective \(\pi\) pulses enable initialization into any spin state. Following a relaxation time \(\tau\), the population in a target state is mapped to \(\ket{0}\) and read out optically. Differences between pairs of the measured relaxation curves yield single-exponential decays from which the rates \(\Omega\) and \(\gamma\) are extracted (Fig.~\ref{fig:introduction}b) \cite{myers2017double, gardill2020fast}. Experiments are conducted using a homebuilt confocal microscope with support for low-temperature (9 K to ambient) and high-temperature (ambient to 474 K) operation modes.
A static magnetic field oriented to within \(5\degree\) of the NV orientation under study is applied with a permanent magnet to lift the \(\ket{\pm 1}\) degeneracy by \(\Delta_{\pm} \approx 145\) MHz. Experiments are conducted using two diamond samples with NV concentrations of 1 ppb (sample A) and \(10^{-3}\) ppb (sample B). 
(See Sec. I in \cite{SupplementalMaterial} for additional experimental details.)

\begin{figure}[t]
\includegraphics[width=0.48\textwidth]{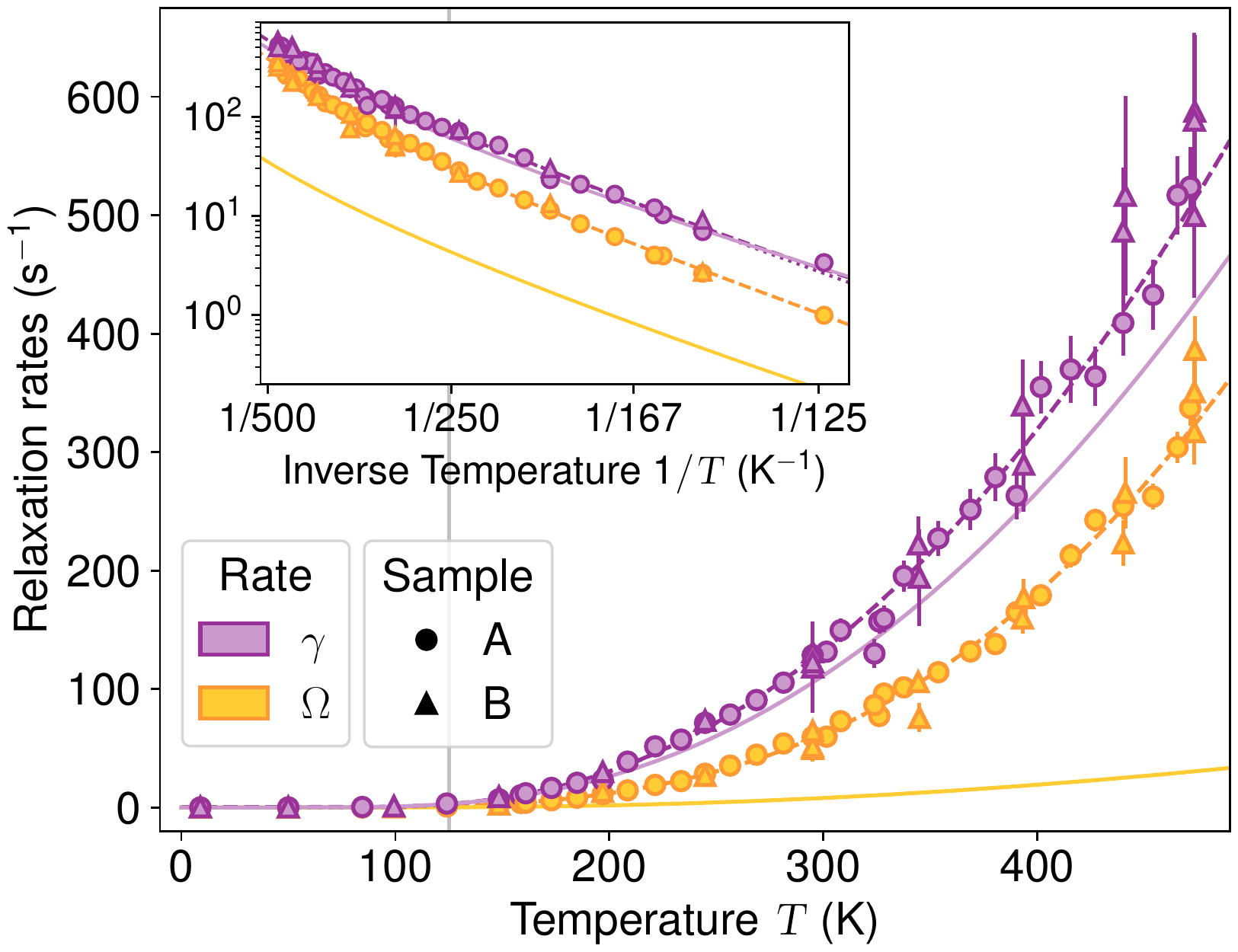}
\caption{\label{fig:rates}
Temperature dependence of relaxation rates \(\gamma\) and \(\Omega\). Error bars are \(1 \sigma\). Above 125 K (vertical gray line), sample-dependent contributions are below 10\%. Darker lines show fit according to the proposed model described by Eqs.~\eqref{double_orbach_omega} and \eqref{double_orbach_gamma}, where dotted (dashed) lines include sample-dependent constants for sample A (B). Lighter solid lines show relaxation rates predicted by \textit{ab initio} calculations. 
See Figs.~S4, S11 and Table~S4 in \cite{SupplementalMaterial} for additional plots and full set of data. Inset: Semi-log plot of relaxation rates versus inverse temperature.
}
\end{figure}

\textit{Results.}---Figure~\ref{fig:rates} displays measurements of the relaxation rates \(\Omega\) and \(\gamma\) as functions of temperature between 9 and 474 K for NV ensemble and single NV measurements in samples A and B respectively. Above 125 K, sample-dependent contributions to \(\Omega\) and \(\gamma\) fall below 10\%, indicating that the relaxation rates are dominated by phonons in the diamond lattice, rather than by interactions with other defects. In Ref.~\cite{cambria2021state}, we found that \(\gamma \approx 2\Omega\) at room temperature. Our results here demonstrate that this factor of 2 is coincidental, as the ratio \(\gamma/\Omega\) declines from 2.5 to 1.8 between 200 and 474 K (Fig. S7 in \cite{SupplementalMaterial}). 

Prior work has fit the temperature dependence of the single-quantum relaxation rate \(\Omega\) using an empirical model in which two terms chiefly contribute to phonon-limited relaxation: one term that scales in proportion to the phonon occupation number at a specific energy, and a second term that scales with temperature as \(T^{5}\). The first term has been described as an ``Orbach'' \cite{redman1991spin}, ``Orbach-type'' \cite{jarmola2012temperature, norambuena2018spin}, or ``Orbach-like'' \cite{cambria2021state} process, as the temperature scaling matches that of the original Orbach process \cite{finn1961spin, orbach1961spin}. Because the NV$^{-}$ center does not exhibit a low-lying excited state, this Orbach-like term has been attributed to quasilocalized phonon modes near meV \cite{norambuena2018spin}. The interactions giving rise to this term have not been described in detail. The second term has been attributed to Raman scattering of low-energy acoustic phonons that are weakly coupled to the spin via first-order interactions, as described by Walker in Ref.~\onlinecite{walker19685}. We argue that Walker's process is negligible in systems like the NV\(^{-}\) ground-state triplet where the spin-phonon interaction strength is much smaller than typical phonon energies. Under this condition, Raman scattering will be driven by second-order, rather than first-order, spin-phonon interactions.

According to Fermi's golden rule, the rate of Raman scattering due to first-order interactions scales quadratically with the squared first-order interaction strength divided by the energy of the virtual state mediating the transition. Gauging the magnitude of the spin-phonon interaction using the zero field splitting, the first-order interaction strength is roughly $h D \cdot (\Delta u/a)$, where $D \approx 2.87$ GHz is the NV zero field splitting, $\Delta u$ is the atomic displacement associated with lattice vibrations, and $a$ is the nearest neighbor distance in diamond. 
If the nearest excited state is beyond the phonon cutoff frequency (i.e. there are no low-lying excited states to enable an Orbach process \cite{finn1961spin, orbach1961spin}),
then the virtual state energy is dominated by the phonon energy, \(\hbar\omega \sim 50\)~meV for acoustic phonons in diamond. Contributions to Raman scattering from second-order interactions scale quadratically with the second-order interaction strength, approximately \(h D \cdot (\Delta u/a)^{2}\). The ratio between the first- and second-order contributions is on the order of \((2 \pi D / \omega)^{2} \sim 10^{-7}\) for the NV center, indicating that Raman scattering via first-order interactions can be neglected. 

Motivated by these observations, we develop an \textit{ab initio} theory of spin-lattice relaxation where Raman scattering is driven by second-order interactions.
The relaxation rate may be expressed \cite{lunghi2022}
\begin{equation}
    \Gamma=\Gamma_{1}^{(1)}(T)+\Gamma_{1}^{(2)}(T)+\Gamma_{2}^{(1)}(T)+\dots\text{,}\label{eq:RelaxTerms}
\end{equation}
where the superscript indicates the order of the spin-phonon interaction (terms with superscript 1 or 2 are linear or quadratic in the atom displacements respectively) and the subscript indicates the order in perturbation theory. 
The term $\Gamma_1^{(1)}$ describes single-phonon processes which are only relevant below 1 K for the NV center~\cite{gugler2018}. At elevated temperatures, relaxation is dominated by Raman scattering of higher energy phonons.

Raman scattering appears as two different second-order terms in Eq.~\ref{eq:RelaxTerms}: first-order interactions applied to second order in perturbation theory \(\Gamma_{2}^{(1)}\) and second-order interactions applied to first order in perturbation theory \(\Gamma_{1}^{(2)}\). 
As discussed above, \(\Gamma_{2}^{(1)} \ll \Gamma_{1}^{(2)}\) for the NV center, and we use Fermi's golden rule to write
\begin{multline}
    \Gamma_{1(m_{s}m_{s}^{\prime})}^{(2)} (T) =
    \frac{2\pi}{\hbar}\sum_{ll^{\prime}}\left|V_{m_{s}m_{s}^{\prime}}^{ll^{\prime}}\right|^{2}\\
    \times \big[n_{l}(n_{l^{\prime}}+1)\delta(\Delta E_{-}) + n_{l^{\prime}}(n_{l}+1)\delta(\Delta E_{+})\big] \label{eq:gamma_21}
\end{multline}
for spin states \(\ket{m_{s}}\) and \(\ket{m_{s}^{\prime}}\). Here, \(V_{m_{s}m_{s}^{\prime}}^{ll^{\prime}}\) is the matrix element coupling \(\ket{m_{s}}\) to \(\ket{m_{s}^{\prime}}\) via phonons \(l\) and \(l^\prime\), and $\Delta E_{\pm} = E_{m_{s}^{\prime}}\pm(\hbar\omega_{l}-\hbar\omega_{l^{\prime}}) - E_{m_{s}}$ is the energy difference between the final and initial states of the composite system. The mean occupation number for mode \(l\) is $n_l = [\mbox{exp}(\hbar \omega_l/k_{\text{B}} T)-1]^{-1}$. 
For high energy phonons, we approximate $\Delta E_{\pm} \approx \pm(\hbar\omega_{l}-\hbar\omega_{l^{\prime}})$ and consider only the diagonal $l=l^\prime$ terms. 
In the continuum limit Eq.~\eqref{eq:gamma_21} becomes
\begin{multline}
\label{eq:gamma_21_continous}
    \Gamma_{1(m_{s}m_{s}^{\prime})}^{(2)} (T) = \frac{4\pi}{\hbar}\intop_{0}^{\infty}d(\hbar\omega)\\
    \times n(\omega)[n(\omega)+1]F_{m_{s}m_{s}^{\prime}}^{(2)}(\hbar\omega,\hbar\omega),
\end{multline}
where the spectral function $F_{m_{s}m_{s}^{\prime}}^{(2)}(\hbar\omega,\hbar\omega)$ accounts for the phonon density of states and the spin-phonon coupling strengths. For the NV center, we identify $\Gamma_{1(\pm0)}^{(2)}$ and $\Gamma_{1(+-)}^{(2)}$ as the relaxation rates $\Omega$ and $\gamma$ respectively. (See Sec. VIII of \cite{SupplementalMaterial} for detailed calculation.)

\begin{figure}[b]
\includegraphics[width=0.48\textwidth]{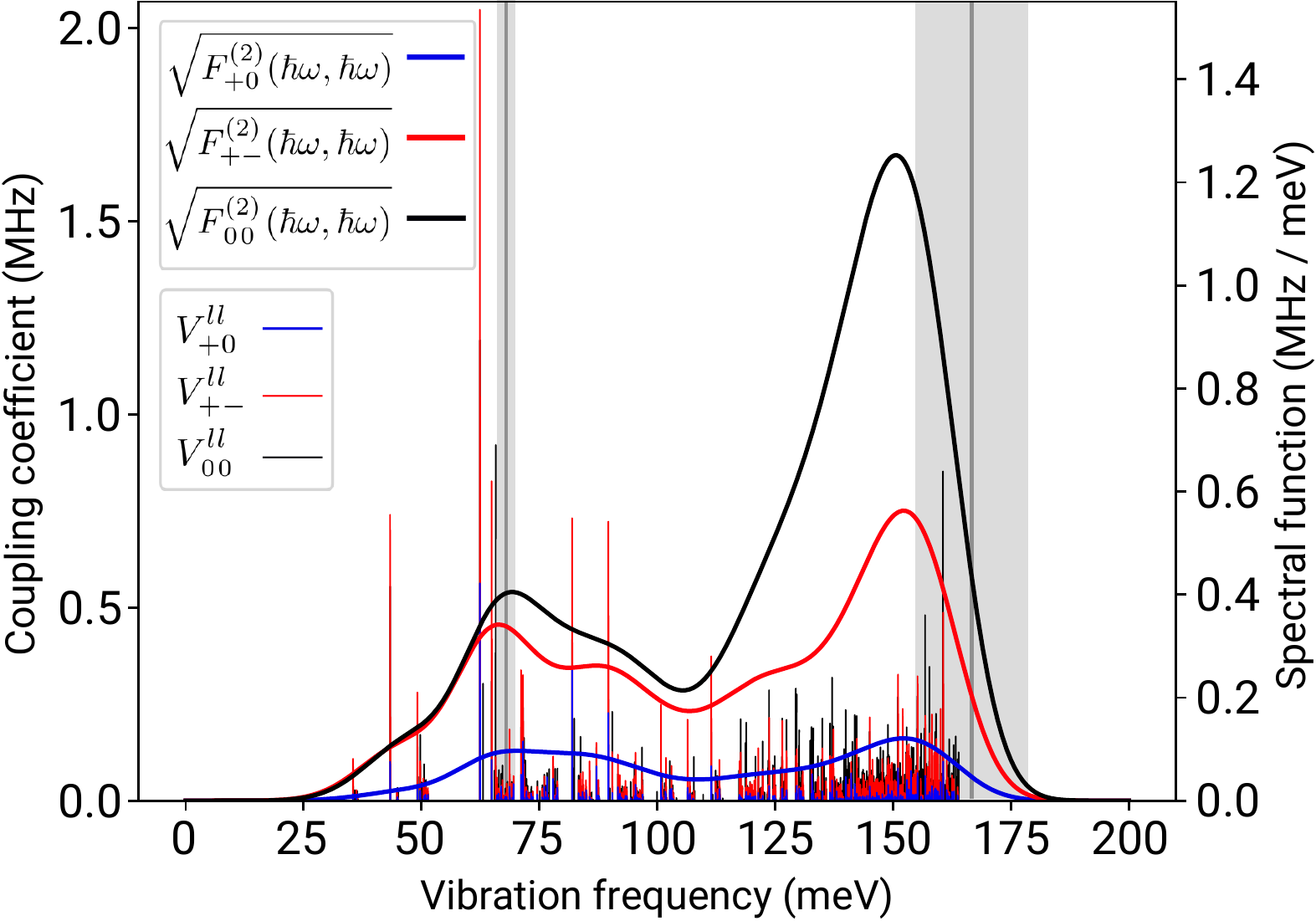}
\caption{\label{fig:simulations} \textit{Ab initio} second-order spin-phonon coupling coefficients (thin lines) and spectral function (thick curves) for a single NV center in a 512 atom supercell. NV spin-phonon dynamics are characterized by the magnitudes of the matrix elements \(\hat{S}_{z}\hat{S}_{+}\) (blue), \(\hat{S}_{+}^{2}\) (red), and \(\hat{S}_{z}^{2} - \tfrac{1}{3}\hat{S}^{2}\) (black), which cause single-quantum relaxation, double-quantum relaxation, and dephasing respectively. The spectral function displays peaks near the values of 68.2(17) and 167(12) meV extracted from the fit of the two-phonon model to the experimental data (gray lines and \(\pm 1 \sigma\) intervals). 
}
\end{figure}

We obtain the matrix elements $V_{m_s m^\prime_s}^{l l}$ by calculating the derivatives of the NV center's spin-spin induced zero field splitting tensor with respect to atomic displacements. 
The calculation is performed using a plane wave supercell density functional theory simulation package~\cite{kresse93, kresse96} by means of PBE functional \cite{func:PBE_96}. 
We approximate the spectral function for a macroscopic diamond by convolving the matrix elements with a normalized Gaussian. The spectral function displays two peaks near 65 and 155~meV (Fig.~\ref{fig:simulations}) which are associated with phonon modes that change the positions of the carbon dangling bonds~\cite{zhang2011vibrational} and thereby the spin density distribution. Evaluating Eq.~\eqref{eq:gamma_21_continous} with the calculated spectral function yields predicted spin-lattice relaxation rates with no free parameters that are in rough quantitative agreement with experiment (lighter solid lines, Fig.~\ref{fig:rates}). Above 125 K, the predicted rates exhibit the same temperature scalings as the measured rates. The double-quantum rate is within 20\% of the measured \(\gamma\), and the single-quantum relaxation rate is approximately 8 times smaller than the measured \(\Omega\) at room temperature (Fig. S9 in \cite{SupplementalMaterial}). The discrepancy for \(\Omega\) may be due to the exclusion of mode combinations where \(l \neq l'\), as combinations of modes with different symmetries likely yield significant matrix elements for differing spin operators, which drive single-quantum transitions. We leave the calculation of off-diagonal combinations for future work. 
The \textit{ab initio} calculations validate the intuitive argument made above for the dominance of second-order interactions in Raman scattering for the NV center. The calculated first-order matrix elements are on the order of 100 MHz, resulting in contributions to relaxation that are roughly 6 orders of magnitude smaller than contributions from second-order matrix elements for 50~meV phonons (Fig. S8 in \cite{SupplementalMaterial}). 

The double-peaked form of the spectral function $F_{m_{s}m_{s}^{\prime}}^{(2)}(\hbar\omega,\hbar\omega)$ suggests that Eq.~\ref{eq:gamma_21_continous} can be well-approximated by an analytical model of the relaxation rates with two terms corresponding to two effective phonon modes:
\begin{gather}
    \Omega(T) = A_{1} \, n_{1} (n_{1} + 1) + A_{2} \, n_{2}(n_{2} + 1) + A_{3}(S),\label{double_orbach_omega}\\
    \gamma(T) = B_{1} \, n_{1} (n_{1} + 1) + B_{2} \, n_{2}(n_{2} + 1) + B_{3}(S).\label{double_orbach_gamma}
\end{gather}
Here, \(n_{1;2} = [\mbox{exp}(\Delta_{1;2}/k_{\text{B}} T)-1]^{-1}\) are the mean occupation numbers at characteristic energies \(\Delta_{1;2}=\hbar \omega_{1;2}\), \(A_{1;2}\) and \(B_{1;2}\) are coupling coefficients for the effective modes, and \(A_{3}(S)\) and \(B_{3}(S)\) are sample-dependent constants. Eqs.~\eqref{double_orbach_omega} and \eqref{double_orbach_gamma} provide excellent fits to the experimental data (darker lines in Fig.~\ref{fig:rates}) with residuals that are consistent with purely statistical errors (Fig. S5 in \cite{SupplementalMaterial}). The characteristic energies from the fit are 68.2(17) and 167(12) meV, matching the locations of the peaks in the spectral function. (See Tables S1, S2 in \cite{SupplementalMaterial} for fit parameters.) Suggestively, the spectral structure of the NV phonon sideband has also been attributed to phonons with the same two characteristic energies as those we extract here~\cite{davies1974vibronic, collins1987nitrogen, alkauskas2014first}. At low temperatures, \(n_{1;2} (n_{1;2} + 1)\approx\exp(-\Delta_{1;2}/k_{\text{B}}T)\). When plotted semi-logarithmically versus inverse temperature, scalings of the form \(n_{1;2} (n_{1;2} + 1)\) therefore appear linear with slope proportional to the activation energy \(\Delta_{1;2}\) (Fig.~\ref{fig:rates} inset). 

\begin{figure}[t]   
\includegraphics[width=0.48\textwidth]{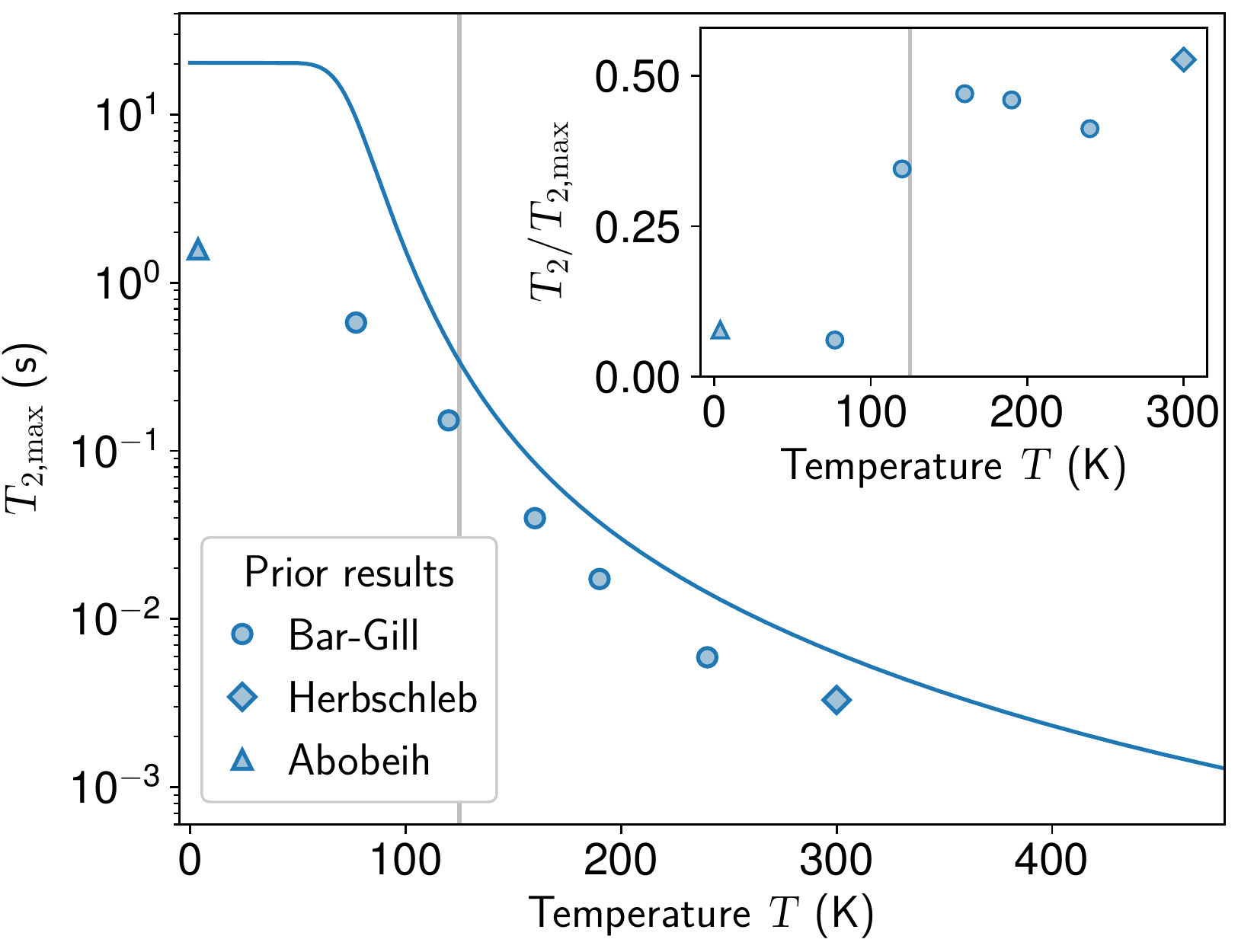}
\caption{\label{fig:T2max}
Temperature dependence of maximum relaxation-limited coherence times. Solid line shows \(T_{2,\text{max}}\) in the single-quantum subspace as a function of temperature according to Eq. \ref{t2_max_qubit}, where \(\Omega\) and \(\gamma\) are calculated from Eqs.~\eqref{double_orbach_omega} and \eqref{double_orbach_gamma} for sample A. (See Fig. S3 in \cite{SupplementalMaterial} for extended plot.) Vertical gray line at 125 K demarcates the phonon-limited regime. Data points show the longest measured NV coherence times at various temperatures as reported in Refs.~\cite{bar2013solid}, \cite{herbschleb2019ultra}, and \cite{abobeih2018one}. Inset: Ratio of measured coherence times to corresponding limits.
}
\end{figure}

The model of NV spin-lattice relaxation proposed here differs qualitatively from the empirical models used in prior works to describe the temperature dependence of the single-quantum relaxation rate \(\Omega\) \cite{redman1991spin, jarmola2012temperature, norambuena2018spin}. Because \(n(n+1) \approx n\) at low temperatures, we suggest that the phenomenological Orbach-like term in prior models corresponds to the first term scaling with \(n_1(n_1+1)\) in Eqs.~\eqref{double_orbach_omega} and \eqref{double_orbach_gamma}. In contrast to prior models, we attribute the high temperature scaling of the spin-lattice relaxation rates to interactions with a second, higher-energy effective phonon mode leading to a second Orbach-like term, rather than to first-order interactions with low-energy acoustic modes leading to a $T^5$ term. Although the prior empirical model can be naively extended to provide a good fit to both measured relaxation rates (Fig. S5 in \cite{SupplementalMaterial}), we argue that it is not physically motivated or necessary to include an additional mechanism with a different scaling, and we predict that the prior model will break down at temperatures beyond those reached in this work.


\textit{Discussion.}---Spin-lattice relaxation is an incoherent process that fundamentally limits achievable coherence times. Much past work with NV centers has neglected the effect of \(\gamma\) on this limit, thereby overestimating the maximum achievable coherence times \cite{bar2013solid,naydenov2011dynamical}. Including \(\gamma\), the relaxation-limited coherence time is
\begin{equation}\label{t2_max_qubit}
    T_{\text{2,max}}^{\text{(SQ)}} = \frac{2}{3\Omega + \gamma}
\end{equation}
for superpositions in the \(\{\ket{0}, \ket{\pm 1}\}\) single-quantum subspace and 
\begin{equation}\label{t2_max_qutrit}
    T_{\text{2,max}}^{\text{(DQ)}} = \frac{1}{\Omega + \gamma}.
\end{equation}
for superpositions in the \(\{\ket{-1}, \ket{+1}\}\) double-quantum subspace \cite{myers2017double, gardill2020fast}. Fig.~\ref{fig:T2max} shows the relaxation limit in the single-quantum subspace alongside the longest measured values of \(T_{2}\) at various temperatures reported in the literature \cite{bar2013solid, herbschleb2019ultra, abobeih2018one}. The best measured coherence times fall short of the calculated limit despite the inclusion of \(\gamma\). The consistency of the ratio between the measured coherence times and \(T_{\text{2,max}}\) in the phonon-limited regime (Fig.~\ref{fig:T2max} inset) suggests that the discrepancy may be due to spin-lattice dephasing, an effect which is not included in Eq.~\ref{t2_max_qubit} \cite{bar2013solid}. The similarity of the spectral function lineshapes for dephasing (Fig.~\ref{fig:simulations}, black curve) and for relaxation (red and blue curves) indicates that these processes may have similar temperature scalings.

Our results have several implications for future research with NV centers and related systems. Because \(T^{5} \gg n(n+1)\) at high temperatures, we predict that NV spin-lattice relaxation will be slower than expected based on prior work at temperatures exceeding those accessed in this or previous studies. The NV spin can be coherently manipulated up to 1000 K \cite{liu2019coherent}, where we predict relaxation rates several times lower than would have been previously expected. Spin-phonon interaction mitigation schemes such as phononic bandgap engineering \cite{cady2019diamond, wolfowicz2021quantum} will likely be ineffective in mitigating the effects of short-wavelength, quasilocalized modes.
On the other hand, it may be possible to reduce relaxation by lowering the occupation number of the 167 meV phonon modes in a manner similar to optical cryocooling \cite{kern2017optical}. In addition, the theoretical model and results presented in this work can be generalized to other crystal defects, such as divacancy centers in silicon carbide~\cite{son2006divacancy, koehl2011nature} and the boron vacancy center in diamond~\cite{umeda2022negatively}, or to other types of spin systems such as molecular qubits, whose spin-lattice dynamics are under active research \cite{escalera2017determining, lunghi2017role, lunghi2022}. In Sec.~IX of \cite{SupplementalMaterial} we discuss the properties of systems that we expect would make them good candidates for extensions of our model.


\textit{Conclusion.}---We have presented measurements of the temperature dependence of the NV center's single- and double-quantum relaxation rates in high-purity diamond samples from 9 to 474 K. We argued that Raman scattering in NV centers primarily arises from second-order spin-phonon interactions, and we have developed widely applicable \textit{ab initio} tools for evaluating the relaxation rates that result from these processes. Our \textit{ab initio} calculation of the NV spin-phonon spectral function demonstrates that two distinct groups of quasilocalized modes provide dominant contributions to NV relaxation at high temperatures. Accordingly, we have developed an analytical model of NV spin-lattice relaxation with just two effective phonon modes with characteristic energies of 68.2(17) and 167(12)~meV. Finally, we calculated the limits imposed by spin-lattice relaxation on the coherence times of superpositions in the single- and double-quantum subspaces, properly accounting for the NV spin triplet over a wide range of temperatures for the first time. 

\vspace{\baselineskip}

We gratefully acknowledge Nathalie de Leon and Maxim Vavilov for helpful conversations and feedback on our manuscript. Experimental work, data analysis, and theoretical efforts conducted at UW--Madison were supported by the U.S.~Department of Energy, Office of Science, Basic Energy Sciences under Award \#DE-SC0020313. Part of this work by V.~L.~was performed under the auspices of the U.S. Department of Energy at Lawrence Livermore National Laboratory under Contract DE-AC52-07NA27344. A.~N.~acknowledges financial support from Fondecyt Iniciaci\'{o}n No 11220266. A.~G.~acknowledges support from the Department of Defense through the National Defense Science and Engineering Graduate Fellowship (NDSEG) program. \'A.~G.~acknowledges the support from the NKFIH in Hungary for the National Excellence Program (Grant No.\ KKP129866), the Quantum Information National Laboratory (Grant No. 2022-2.1.1-NL-2022-00004), and the EU QuantERA II MAESTRO project and from the European Commission for the QuMicro project (Grant No.\ 101046911). G.~T.~was supported by the János Bolyai Research Scholarship of the Hungarian Academy of Sciences. \'A.~G.~and G.~T.~acknowledge the high-performance computational resources provided by KIFÜ (Governmental Agency for IT Development) institute of Hungary. J.~R.~M.~acknowledges support from ANID-Fondecyt 1221512. 

\bibliography{MainReferences.bib}

\end{document}


\preprint{APS/123-QED}

\title{Supplemental Material for ``Temperature-dependent spin-lattice relaxation of the nitrogen-vacancy spin triplet in diamond''}

\author{M.~C.~Cambria}
 
\affiliation{Department of Physics, University of Wisconsin, Madison, Wisconsin 53706, USA}

\author{A.~Norambuena}

\affiliation{Centro de Optica e Informaci\'on Cu\'antica, Universidad Mayor, Camino La Pir\'amide 5750, Huechuraba, Santiago, Chile}

\author{H.~T.~Dinani}
 
\affiliation{Centro de Investigaci\'on DAiTA Lab, Facultad de Estudios Interdisciplinarios, Universidad Mayor, Santiago, Chile}

\affiliation{Escuela Data Science, Facultad de Ciencias, Ingenier\'{i}a  y Tecnolog\'{i}a, Universidad Mayor, Santiago, Chile}

\author{G.~Thiering}

\affiliation{Wigner Research Centre for Physics, P.O. Box 49, 1525 Budapest, Hungary}

\author{A.~Gardill}

\author{I.~Kemeny}

\author{Y.~Li}
 
\affiliation{Department of Physics, University of Wisconsin, Madison, Wisconsin 53706, USA}

\author{V.~Lordi}
 
\affiliation{Lawrence Livermore National Laboratory, Livermore, CA, 94551, USA}

\author{\'A.~Gali}

\affiliation{Wigner Research Centre for Physics, P.O. Box 49, 1525 Budapest, Hungary}

\affiliation{Department of Atomic Physics, Institute of Physics, Budapest University of Technology and Economics, M\H{u}egyetem rakpart 3., 1111 Budapest, Hungary}

\author{J.~R.~Maze}
 
\affiliation{Instituto de F\'isica, Pontificia Universidad Cat\'olica de Chile, Casilla 306, Santiago, Chile}

\affiliation{Centro de Investigaci\'on en Nanotecnolog\'ia y Materiales Avanzados, Pontificia Universidad Cat\'olica de Chile, Santiago, Chile}

\author{S.~Kolkowitz}
\email{kolkowitz@wisc.edu}
 
\affiliation{Department of Physics, University of Wisconsin, Madison, Wisconsin 53706, USA}

\maketitle

\section{Experimental details}

\begin{figure}[t]
\includegraphics[width=0.48\textwidth]{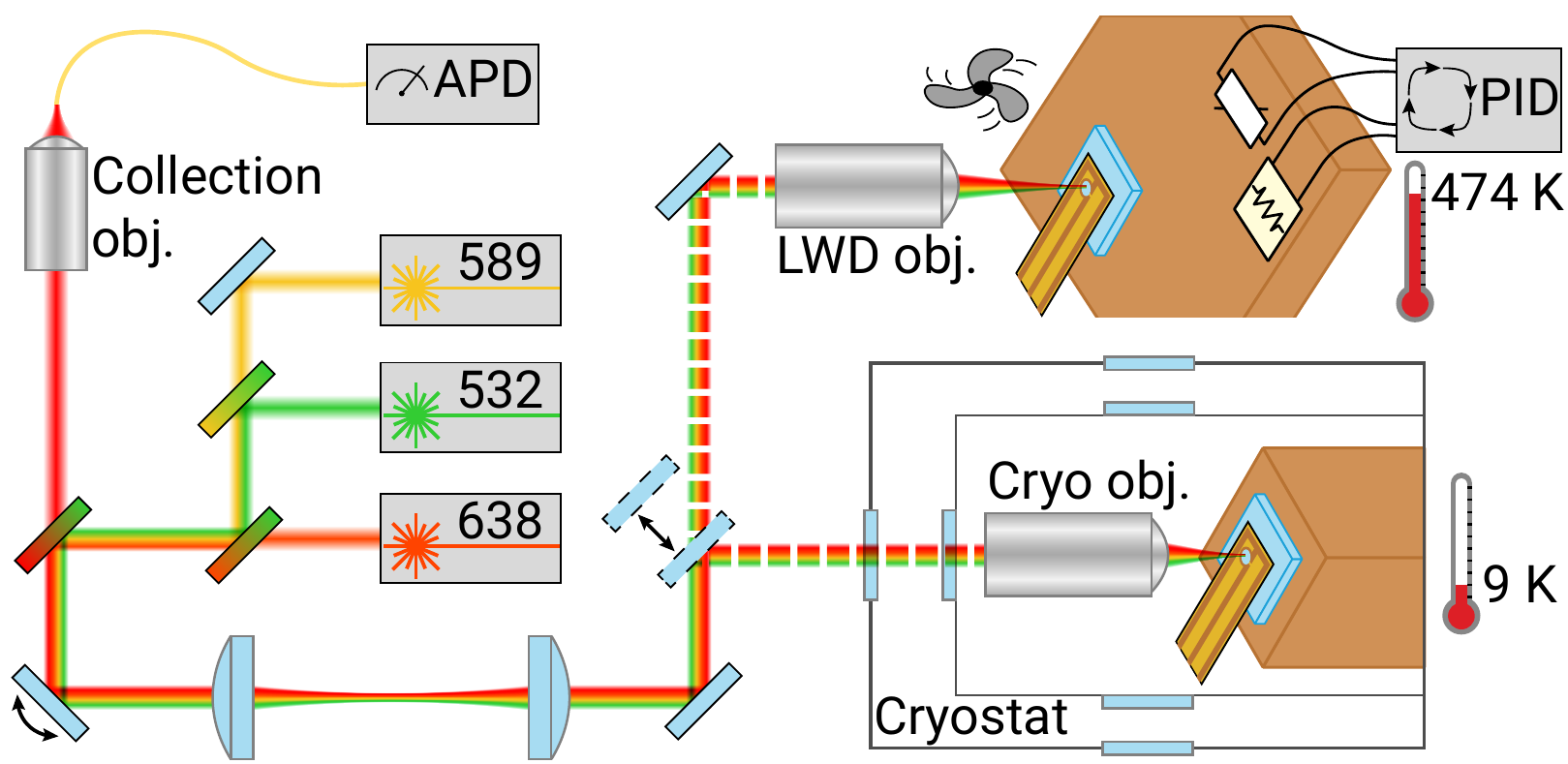}
\caption{\label{fig:apparatus} 
Diagram of the experimental apparatus supporting spin-to-charge-conversion readout and dual low- and high-temperature operation modes. 
}
\end{figure}

Experiments were conducted using a homebuilt confocal microscope with dual support for low-temperature (9 K to ambient) and high-temperature (ambient to 474 K) operation modes (Fig.~\ref{fig:apparatus}). Switching between modes is accomplished with a flip-mounted mirror that routes the optical path between the low- and high-temperature setups. The low-temperature setup consists of an attocube attoDRY800 closed-cycle cryostat with a 0.82 NA cryogenic objective (attocube LT-APO/VISIR/0.82). The high-temperature setup consists of a homebuilt copper hot plate with a 0.55 NA long working distance (LWD) objective (Edmund Optics 59-879). The front of the LWD objective is actively cooled by air convection from a 12 V \(50\times50\times10\) mm case fan. In order to minimize vibrations, the case fan is mounted on a bench separate from the optical table and airflow is directed to the objective via a hose attached to the fan. An apertured aluminum plate is used to isolate the hot plate from the airflow. In our experience, an objective which has a shorter working distance or is not actively cooled can reach high temperatures and sustain permanent damage as a result. The temperature of the hot plate is stabilized by a custom proportional-integral-derivative (PID) feedback loop. In both operation modes, the diamond sample is thermally anchored to the cold finger/hot plate using conductive silver paint (Ted Pella Leitsilber). 
To accurately report the temperature of the diamond sample in high-temperature operation mode, we use the NV as a temperature sensor by inverting the temperature dependence of the zero field splitting reported in Ref.~\cite{toyli2012measurement}. For low-temperature operation mode, we calibrated the cryostat's integrated sample temperature sensor against an additional temperature sensor (Lakeshore CX-1050-SD-HT-1.4L-QT) mounted in the same manner as a diamond sample. 

As stated in the main text, the experimental sequence used to extract the relaxation rates is the same as that used in prior works \cite{myers2017double, gardill2020fast, cambria2021state}. Briefly, the sequence consists of the following steps: (1) optical polarization into \(m_{s}=0\), (2) a state-selective \(\pi\) pulse to transfer population into a target initialization spin state, (3) some relaxation time \(\tau\), (4) a state-selective \(\pi\) pulse to transfer population from a target readout spin state to \(m_{s}=0\), and (5) optical readout of the population in \(m_{s}=0\). This procedure is repeated for different values of \(\tau\) and initialization/readout states, and the relaxation rates are extracted from the resultant population curves. Further details are available in the above-mentioned works \cite{myers2017double, gardill2020fast, cambria2021state}.

Optical polarization is achieved using a 1 ms pulse of approximately 1.5 mW of 532 nm light.The population in \(\ket{0}\) is readout optically by collecting fluorescence from the phonon sideband above 638 nm. Microwaves are delivered using a coplanar waveguide (CPW) mounted on top of the diamond sample. The CPW has an \(\Omega\)-shaped loop along the signal trace which is shorted to the ground trace. The loop features an 1 mm diameter aperture in the center to allow for optical access to the sample underneath. It was manufactured by PCBWay and consists of a single 0.5 oz gold-plated copper layer on a 25 micron polyimide substrate. A coaxial connection is provided by a coplanar MMCX end-launch soldered onto pads at the end of the CPW.

Data from two different chemical-vapor-deposition grown samples is presented. The NVs in both samples are formed naturally during diamond growth. Sample A, from Chenguang Machinery \& Electric Equipment Company, displays an NV concentration of approximately 1 ppb. Measurements in this sample were performed on the ensemble of NVs within a single confocal volume, which we estimate contains about 30 (150) NVs of a given orientation for low-temperature (high-temperature) operation mode. Ensembles in high-temperature operation mode contain more NVs than ensembles in low-temperature operation mode due to the difference in NA between the low- and high-temperature mode objectives. 
Sample B, from Element Six, displays an inhomogeneous NV concentration with different areas exhibiting concentrations between \(10^{-5}\) and \(10^{-3}\) ppb. The higher-concentration regions are localized in irregularly-spaced planes, each approximately 5 microns thick. Measurements in sample B were conducted with single NVs isolated within the higher-concentration planes.
The NV concentrations quoted in the main text are estimated by NV counting or by comparing the observed fluorescence rate to that of a single isolated NV. In both samples, measurements were conducted with NVs several microns below the diamond surface so as to avoid the effects of surface noise. 

In several instances, we use a modified version of the standard relaxation rate experiment sequence described in prior works \cite{myers2017double, gardill2020fast, cambria2021state} in order to achieve more efficient measurements. For all measurements in sample A (the higher NV concentration sample), we use Knill composite \(\pi\) pulses in place of standard \(\pi\) pulses to drive spin state transitions. Knill pulses consist of five standard \(\pi\) pulses with differing phases (30\degree, 0\degree, 90\degree, 0\degree, 30\degree) applied in series. Knill pulses, like other composite \(\pi\) pulses, are more robust against off-resonance errors than standard \(\pi\) pulses and so help to mitigate the effect of inhomogeneous broadening within ensembles \cite{ryan2010robust, souza2011robust}. For measurements in sample B (the lower NV concentration sample) below room temperature, we use spin-to-charge-conversion (SCC) readout in place of standard optical readout. In the standard readout scheme, NVs are excited with green illumination (\({\sim}1.5\) mW) and the resultant spin-state dependent fluorescence is collected. Because the spin repolarizes during this process, readout is limited to roughly 300 ns during which \(<1\) photon is collected on average per experiment for single NVs. Each experiment must therefore be repeated \(>10^{5}\) times for sufficient averaging, leading to prohibitively long single-NV measurements at low temperatures. In contrast, in SCC readout a spin-state selective optical ionization pulse (\({\sim}30\) mW for \({\sim}100\) ns at 638 nm) maps the NV spin state onto its charge state, which is then readout via fluorescence under weak (\({\sim}10\) \(\upmu\)W) yellow illumination \cite{shields2015efficient}. The charge state is robust under the readout illumination, enabling long readouts (\({\sim}10\) ms) during which tens of photons may be collected. For long experiments where the overhead of SCC readout is relatively small, SCC readout results in an enhanced measurement sensitivity that allows for the practical measurement of slow relaxation rates with single NVs. We note that unlike in the original description of SCC readout \cite{shields2015efficient}, we do not use a shelving pulse, as we found the shelving pulse to have no discernible effect on readout fidelity during characterization. In order to verify that the modifications discussed in this paragraph (Knill pulses and SCC readout) do not introduce errors or artifacts into our measurements, we ran relaxation experiments at room temperature with standard and SCC readout as well as with standard \(\pi\) pulses and Knill \(\pi\) pulses and confirmed in each case that the modified sequences yield the same relaxation rates.

The assumption of a single-valued \(\Omega\) to describe phonon-limited relaxation on both the \(\ket{0} \leftrightarrow \ket{-1}\) and \(\ket{0} \leftrightarrow \ket{+1}\) transitions has been verified empirically in previous work \cite{cambria2021state} and is consistent with all experimental data collected for this study.

\section{Fit parameters for the proposed model}

The fit parameters for the proposed model in the phonon-limited regime are presented in Table~\ref{tab:double_orbach}. The sample-dependent constants extracted from the same fit are presented separately in Table~\ref{tab:sample_dependent_constants}. 

\begin{table*}[t]
    \begin{tabularx}{0.8\textwidth}{|>{\centering\arraybackslash}X >{\centering\arraybackslash}X >{\centering\arraybackslash}X >{\centering\arraybackslash}X >{\centering\arraybackslash}X >{\centering\arraybackslash}X |} 
     \hline
     \(A_{1}\) (s\(^{-1}\)) & \(B_{1}\) (s\(^{-1}\)) & \(\Delta_{1}\) (meV) & \(A_{2}\) (s\(^{-1}\)) & \(B_{2}\) (s\(^{-1}\)) & \(\Delta_{2}\) (meV) \\
     \hline
     \num{0.58(8)e3} & \num{1.51(17)e3} & 68.2(17) & \num{9(2)e3} & \num{4.8(14)e3} & 167(12) \\
     \hline
    \end{tabularx}
    \caption{Fit parameters for \(\Omega\) and \(\gamma\) according to the proposed model described by Eqs. (4) and (5) of the main text. Uncertainties are \(1\sigma\).}
    \label{tab:double_orbach}
\end{table*}

\section{Comparison to prior measurements}

\begin{figure}[b]
\includegraphics[width=0.98\textwidth]{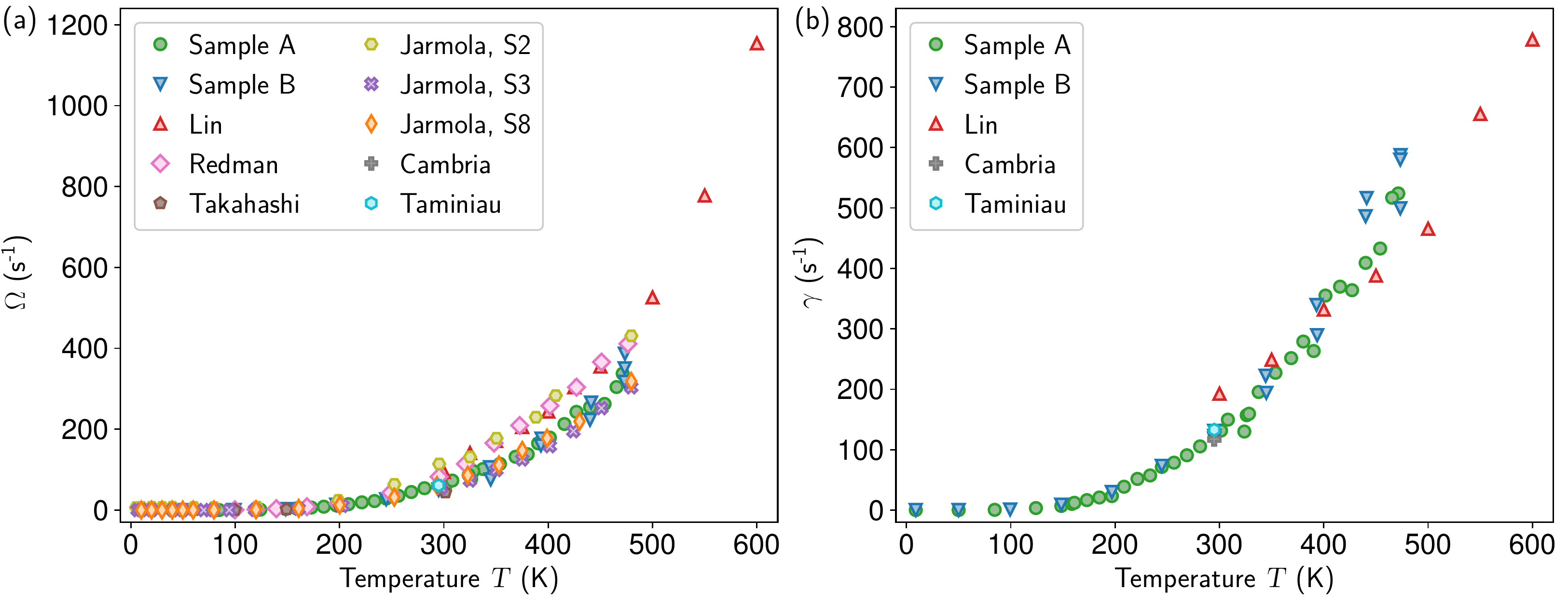}
\caption{\label{fig:prior_work_comparison}
Comparison of prior experimental measurements of the single- and double-quantum relaxation rates (panels a and b respectively). Data is reproduced from Refs.~\cite{redman1991spin, takahashi2008quenching, jarmola2012temperature, taminiau2014universal, cambria2021state, lin2021temperature}. 
}
\end{figure}

Several experimental studies have measured the temperature dependence of the single-quantum relaxation rate in bulk diamond samples  \cite{redman1991spin, takahashi2008quenching, jarmola2012temperature}. In a recent study \cite{lin2021temperature}, Lin et al. report measurements of the temperature dependence of both the single- and double-quantum relaxation rates above room temperature. Measurements of the single- and double-quantum rates are reported in Refs. \cite{taminiau2014universal} and \cite{cambria2021state}, but only at room temperature. The results from these prior studies are shown alongside the results presented in this work in Fig.~\ref{fig:prior_work_comparison}. Our measurements of the single-quantum relaxation rate are consistent with the majority of prior results in low-density samples (\([\text{NV}] < 1 \text{ ppm}\)) to within approximately 10\% \cite{takahashi2008quenching, jarmola2012temperature, taminiau2014universal, cambria2021state}. Interestingly, in the high-density samples (\([\text{NV}] \sim 10 \text{ ppm}\)) measured by Redman et al. and Jarmola et al. (sample S2), single-quantum relaxation rates were approximately 30\% faster than the typical values seen in low-density samples \cite{redman1991spin, jarmola2012temperature}. Jarmola proposed that this discrepancy is due to temperature-dependent cross-relaxation, an effect which is not captured by the sample-dependent constant terms in the model described by Eqs.~(4) and (5) of the main text. 

The single-quantum relaxation rates measured by Lin et al. in Ref.~\cite{lin2021temperature} are consistent with the rates measured by Redman and Jarmola in high-density samples, but surprisingly, the measurements in Ref.~\cite{lin2021temperature} were conducted using a low-density sample with a reported NV concentration of only 2 ppb. Lin's observation of fast single-quantum relaxation in a low-density sample may indicate that the temperature-dependent cross-relaxation proposed by Jarmola involves a separate defect species; in this case NV concentration would not be a reliable indicator of the magnitude of cross-relaxation. In addition, Lin's measurement of the double-quantum relaxation rate at room temperature for low-density samples is around 70\% faster than the value established by previous work \cite{taminiau2014universal, cambria2021state} and replicated in this work. The temperature dependence of the double-quantum relaxation rate reported by Lin is also qualitatively different than that reported in this work; in Lin's measurements, the double-quantum relaxation rate increases with temperature more slowly than the single-quantum relaxation, such that \(\Omega > \gamma\) at \(T > 450 \text{ K}\). In contrast we find \(\Omega < \gamma\) at all temperatures in the phonon-limited regime. It is difficult to reconcile this observation with our findings, even allowing for a temperature-dependent cross relaxation process involving a high concentration of dark defects in Lin's sample. We conclude that the results of Ref.~\cite{lin2021temperature} are not entirely unprecedented, but are probably not applicable to the vast majority of samples studied by the NV community. 

\begin{figure}[t]
\includegraphics[width=0.48\textwidth]{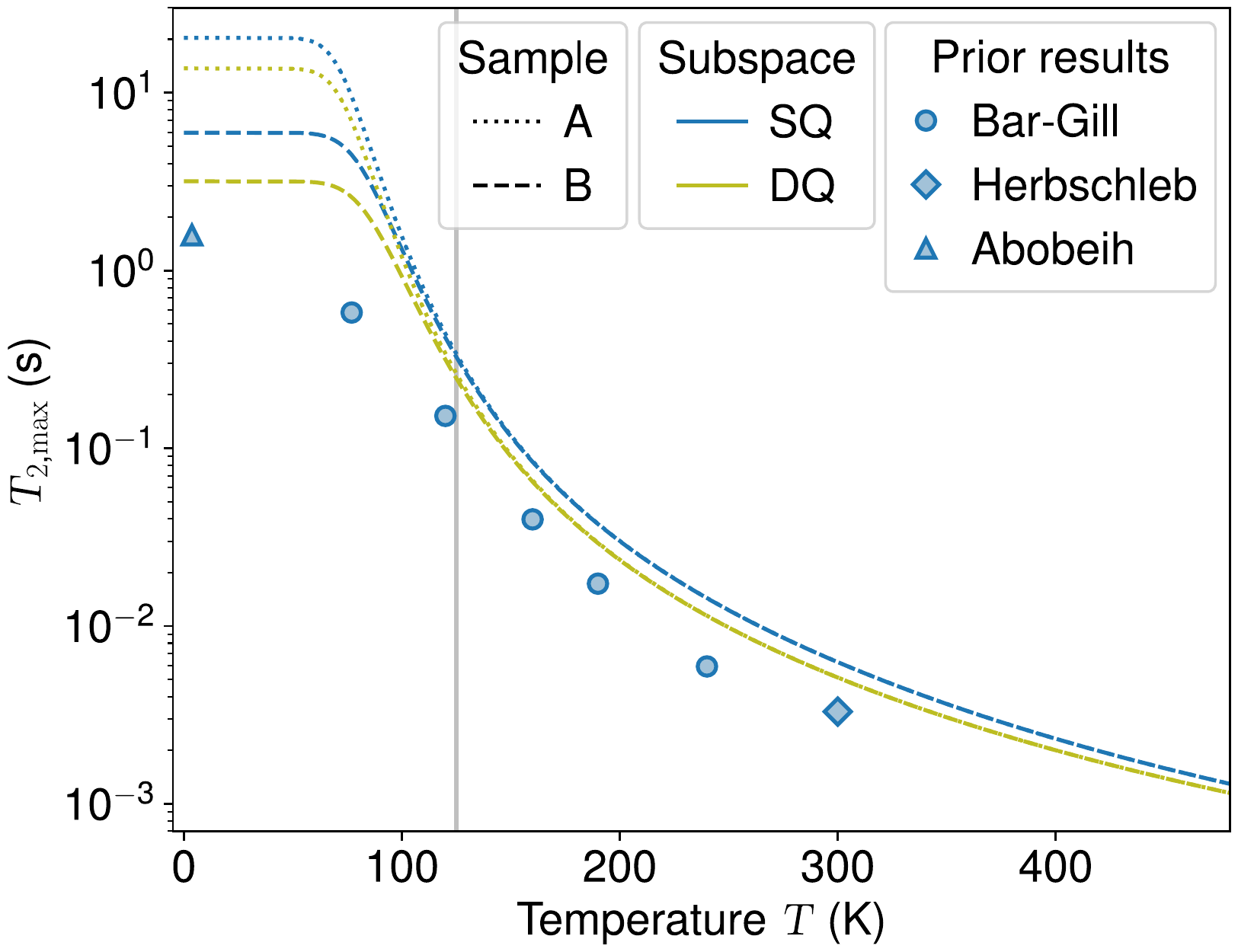}
\caption{\label{fig:t2max-extended}
Extended version of Fig.~4 from the main text. Sample B (dashed lines) exhibits shorter maximum relaxation-limited coherence times than sample A (dotted lines) due to the high sample-dependent constant for the double-quantum rate \(\gamma\) in Sample B. As in the main text version of this figure, the data points show the longest experimentally measured coherence times in the single-quantum subspace at various temperatures as reported in Refs.~\cite{bar2013solid, herbschleb2019ultra, abobeih2018one}. 
}
\end{figure}

\section{Sample-dependent behavior at low temperatures}

\begin{figure}[b]
\includegraphics[width=0.98\textwidth]{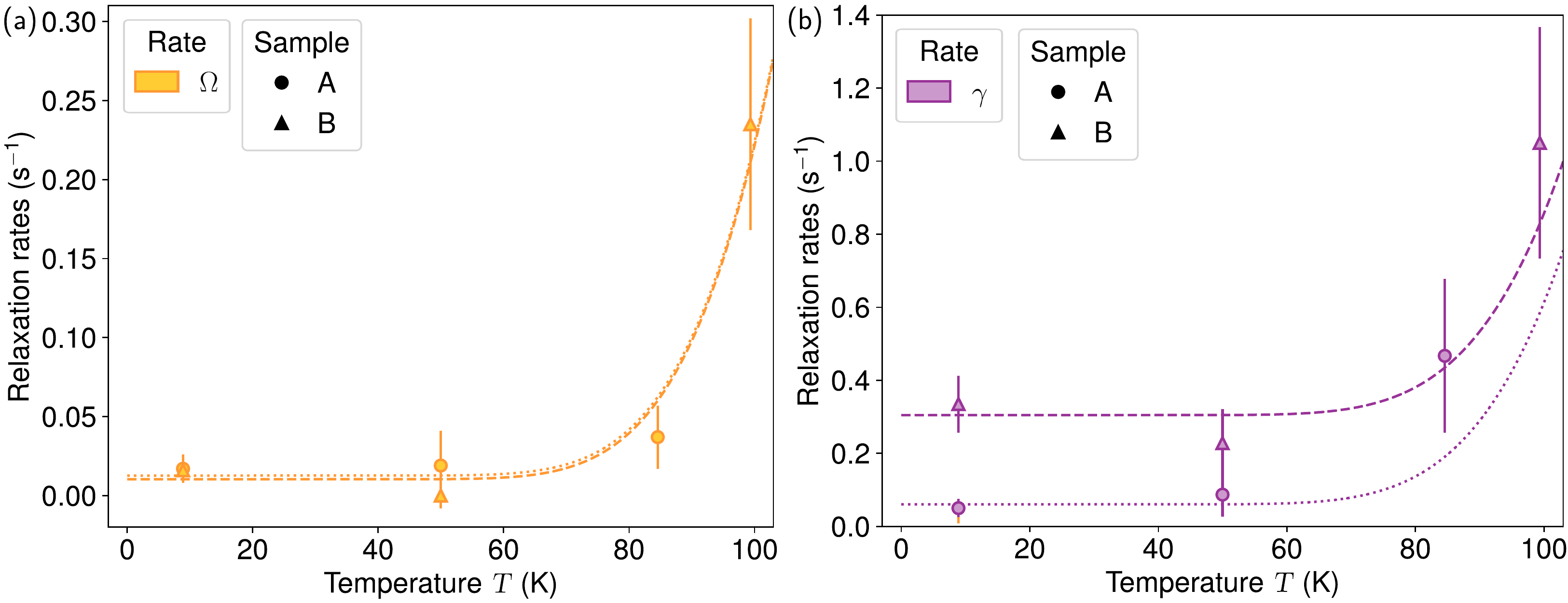}
\caption{\label{fig:sample_dependent_zoom-extended}
Detail views of sample-dependent behavior at low temperatures for the single- and double-quantum relaxation rates (panels a and b respectively).
}
\end{figure}

At temperatures below 125 K, sample-dependent constants begin to make non-negligible contributions to the relaxation rates. The values of these constants as determined from the fit of the proposed model (Eqs.~(4) and (5) in the main text) to the experimental data are shown in Table~\ref{tab:sample_dependent_constants}. We note that sample-dependent constants have been attributed to interactions between defects \cite{jarmola2012temperature, norambuena2018spin}. Accordingly, it may be expected that samples with higher defect concentrations exhibit larger sample-dependent constants. The samples studied for this work do not display a significant difference between their respective constants for \(\Omega\). Interestingly, we observe that the constant for \(\gamma\) is around 5 times larger in sample B (\([\text{NV}] \approx 10^{-3} \text{ ppb}\)) than in sample A (\([\text{NV}] \approx 1 \text{ ppb}\)). While a full investigation of this effect is out of scope for the present work, we suggest that the unexpectedly large constant for \(\gamma\) in sample B may be associated with a dark defect species present at higher levels in sample B than in sample A. Fig.~\ref{fig:t2max-extended} shows the effect of the sample-dependent constants on the relaxation-limited coherence times for samples A and B. We see that coherence times in sample A can in principle be over three times longer than in sample B at low temperatures as a result of sample-dependent relaxation.

\begin{table*}[t]
    \begin{tabularx}{0.9\textwidth}{|>{\centering\arraybackslash}X >{\centering\arraybackslash}X >{\centering\arraybackslash}X >{\centering\arraybackslash}X |} 
     \hline
     \(A_{3} \ (\text{Sample A})\) (s\(^{-1}\)) & \(B_{3} \ (\text{Sample A})\) (s\(^{-1}\)) & \(A_{3} \ (\text{Sample B})\) (s\(^{-1}\)) & \(B_{3} \ (\text{Sample B})\) (s\(^{-1}\)) \\
     \hline
     0.013(8) & 0.06(2) & 0.010(5) & 0.30(6) \\
     \hline
    \end{tabularx}
    \caption{Sample-dependent constants from the fit of the model described by Eqs.~(4) and (5) of the main text to the experimental data. The sample-dependent constant for \(\Omega\) (\(\gamma\)) is denoted by \(A_{3}\) (\(B_{3}\)).}
    \label{tab:sample_dependent_constants}
\end{table*}

\begin{figure}[b]
\includegraphics[width=0.48\textwidth]{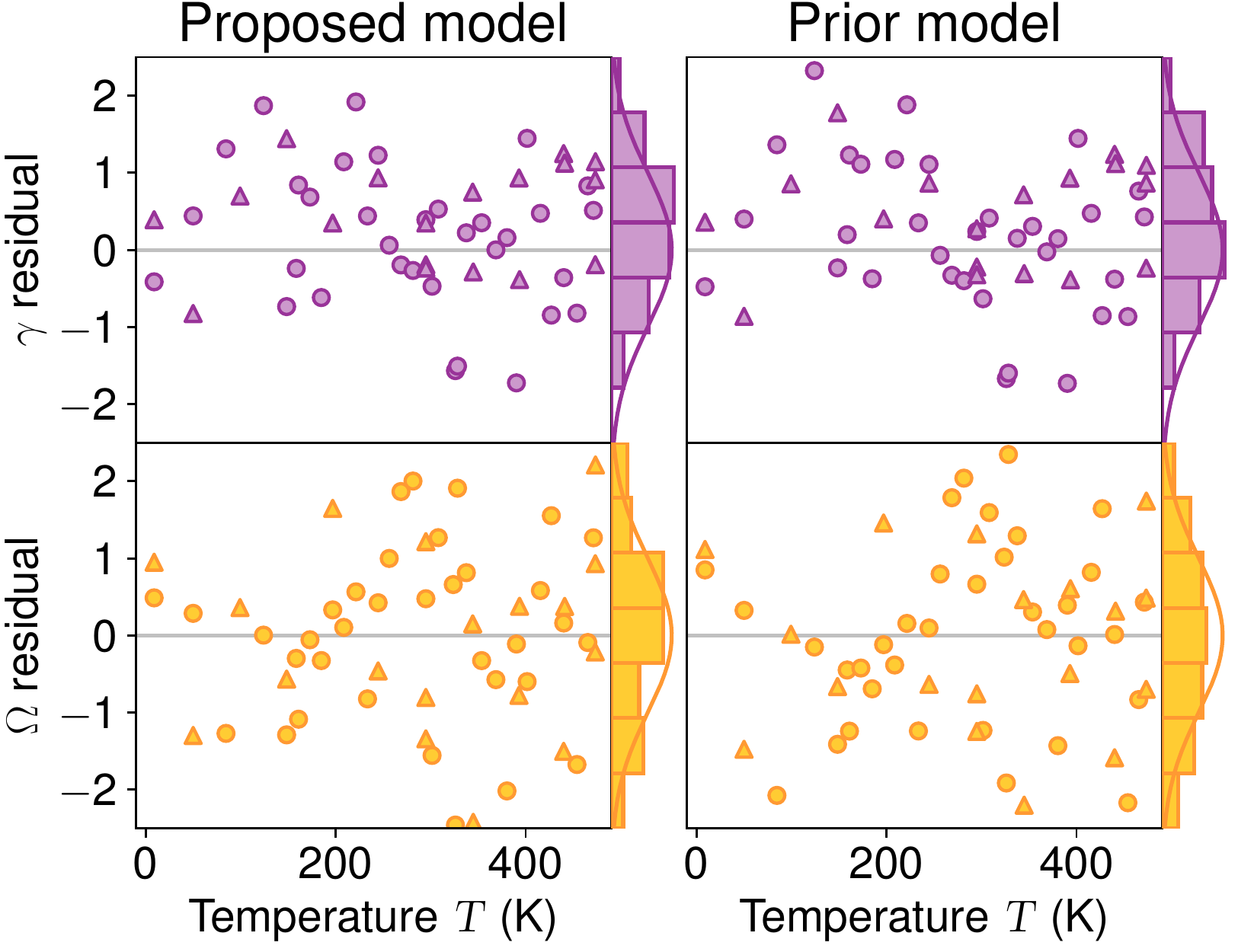}
\caption{\label{fig:residuals}
Normalized residuals for the proposed model (Eqs.~(4) and (5) of the main text) and the prior model (Eqs.~\eqref{orbach_t5_omega} and \eqref{orbach_t5_gamma} of this document). For clarity, the extent of each plot is limited to \(\pm 2.5 \sigma\); some outliers are not visible. Histograms of the residuals shown at the edges of the plots are consistent with a normal distribution of mean 0 and variance 1 (solid lines), indicating both models provide excellent fits to the data.
}
\end{figure}

\section{Comparison between the proposed and prior analytical models}

In prior work \cite{redman1991spin, jarmola2012temperature, norambuena2018spin}, the analytical model used to describe the temperature dependence of the single-quantum relaxation rate \(\Omega\) typically contains terms accounting for low-temperature behavior, an Orbach-like term which scales with the occupation number \(n = [\mbox{exp}(\Delta/k_{\text{B}} T)-1]^{-1}\) at a phonon energy \(\Delta\), and a \(T^{5}\) term attributed to the effect of low-energy weakly coupled acoustic phonons. Here we develop an extended version of this prior model so that we can compare it to the model we propose in the main text (Eqs.~(4) and (5)). As discussed in the main text, prior experimental work has tacitly assumed \(\gamma=0\) \cite{redman1991spin, takahashi2008quenching, jarmola2012temperature}. In these contexts the lifetime of \(\ket{0}\) is denoted \(T_{1}\), which is equivalent to \(1/3\Omega\). For clarity we denote this quantity \(T_{1}^{(0)}\) here. As a starting point, we consider the expression used by Jarmola et al. in Ref.~\cite{jarmola2012temperature}:
\begin{gather}
    \frac{1}{T_{1}^{(0)}} = A_{1}(S) + \frac{A_{2}}{e^{\Delta/k_{\text{B}}T} - 1} + A_{3} \, T^{5}.\label{jarmola_omega}
\end{gather}
We note that the occupation number \(n\) is written in exponential form. As the Orbach-like term describes scattering of phonons at a specific energy \(\Delta\), the full temperature dependence of this term is \(n(n+1)\) \cite{cambria2021state}. Jarmola et al. simplify this by taking the low-temperature approximation \(n(n+1) \approx n\) for \(n \ll 1\); we avoid this approximation here. With these considerations, we extend Eq.~\eqref{jarmola_omega} to describe both the single- and double-quantum relaxation rates \(\Omega\) and \(\gamma\):
\begin{gather}
    \Omega(T) = A_{1} \, n (n + 1) + A_{2} \, T^{5} + A_{3}(S),\label{orbach_t5_omega}\\
    \gamma(T) = B_{1} \, n (n + 1) + B_{2} \, T^{5} + B_{3}(S).\label{orbach_t5_gamma}
\end{gather}
Fits to the experimental data according to Eqs.~\eqref{orbach_t5_omega} and \eqref{orbach_t5_gamma} are nearly identical to those provided by the proposed model described by Eqs.~(4) and (5) of the main text and shown in Fig.~2 of the main text. The normalized residuals for the two models are shown in Fig.~\ref{fig:residuals}. We see that both the proposed model (Eqs.~(4) and (5) of the main text) and the prior model (Eqs.~\eqref{orbach_t5_omega} and \eqref{orbach_t5_gamma}) provide excellent fits to the experimental data. Quantitatively, the reduced chi-squared metric for the proposed model is \(\chi_{\nu}^{2}=1.25\), slightly better than that for the prior model, \(\chi_{\nu}^{2}=1.29\). As such, in the temperature range studied in this work it is not possible to say which model is more physically accurate from the fits alone. However, the terms \(T^{5}\) and \(n(n+1)\) exhibit dramatically different scalings at higher temperatures than those accessed in this or prior studies. Accordingly, it may be possible to demonstrate the breakdown of the prior model at temperatures of around 700 K, where the prior model predicts relaxation rates around 50\% (20\%) faster than the prior model for \(\Omega\) (\(\gamma\)).

\section{Comparison between versions of the proposed model with different numbers of effective phonon modes}

\begin{figure}[t]
\includegraphics[width=0.98\textwidth]{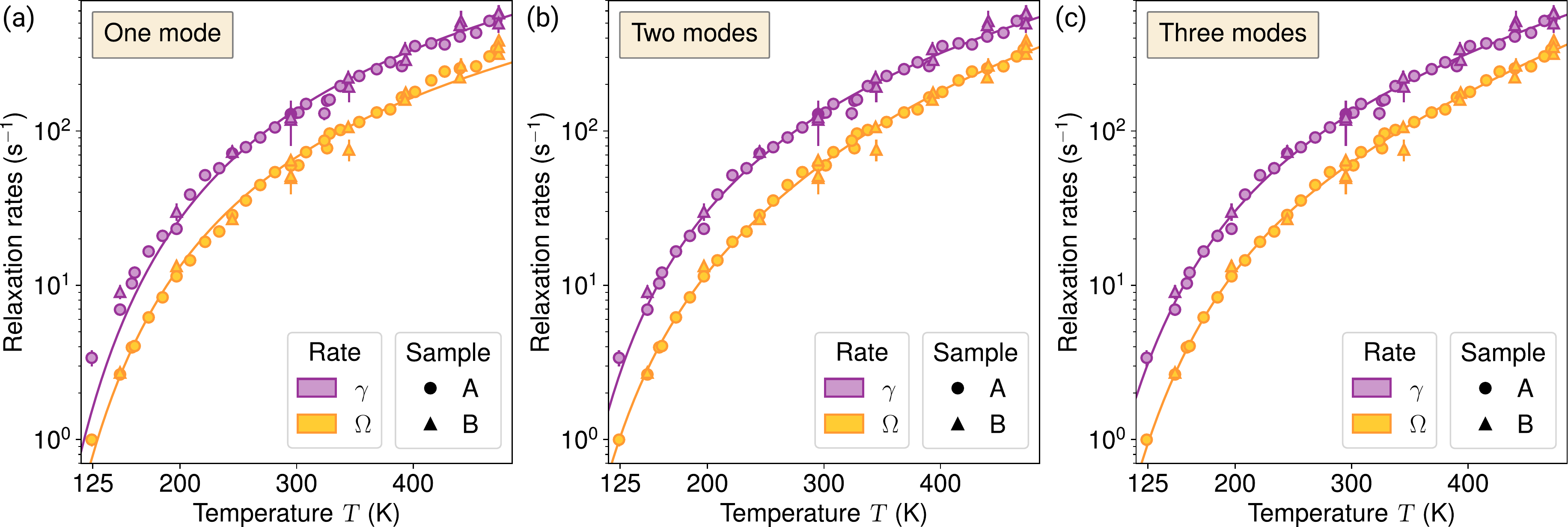}
\caption{\label{fig:n_orbach}
Semi-log plots of fits to the experimental data for versions of the proposed model with one, two, and three modes (panels a, b, and c respectively). For clarity, the sample-dependent constants are removed from the plotted functions after the fits are performed and the plots are restricted to the phonon-limited regime above 125 K.
}
\end{figure}

The analytical model we propose (Eqs. 4 and 5 of the main text) assumes that the NV spin-phonon spectral function is best approximated by a sum of two strongly coupled effective phonon modes. It is sensible to ask whether a similar model consisting of just one mode provides an adequate fit to the experimental data, or if a model with more modes provides a superior fit. Fig.~\ref{fig:n_orbach} demonstrates the fits provided by models consisting of one, two, and three effective phonon modes. We see that the the one-mode model (panel a) does not provide an adequate fit to the data, as reflected by its poor reduced chi-squared statistic of \(\chi_{\nu}^{2}=3.88\). The one-mode model identifies an activation energy of 80.5(6) meV, which lies between the energies identified by the two-mode model, 68.2(17) and 167(12)~meV. The two-mode model provides an excellent fit to the data (panel b), with a reduced chi-squared of \(\chi_{\nu}^{2}=1.32\). Adding a third mode does not improve the fit in a clear visual way (panel c), and improves the reduced chi squared only marginally to \(\chi_{\nu}^{2}=1.25\). While the activation energies in the two-mode version of the model are consistent with our own \textit{ab initio} calculations and prior spectroscopic results, the activation energies in the three-mode version of the model less well-motivated, at 55(14) meV, 84(13) meV, and \(2.5(5) \times 10^{2}\) meV. We note that the highest activation energy here is nonphysical, lying well above the maximum phonon energies in diamond.

\begin{figure}[t]
\includegraphics[width=0.48\textwidth]{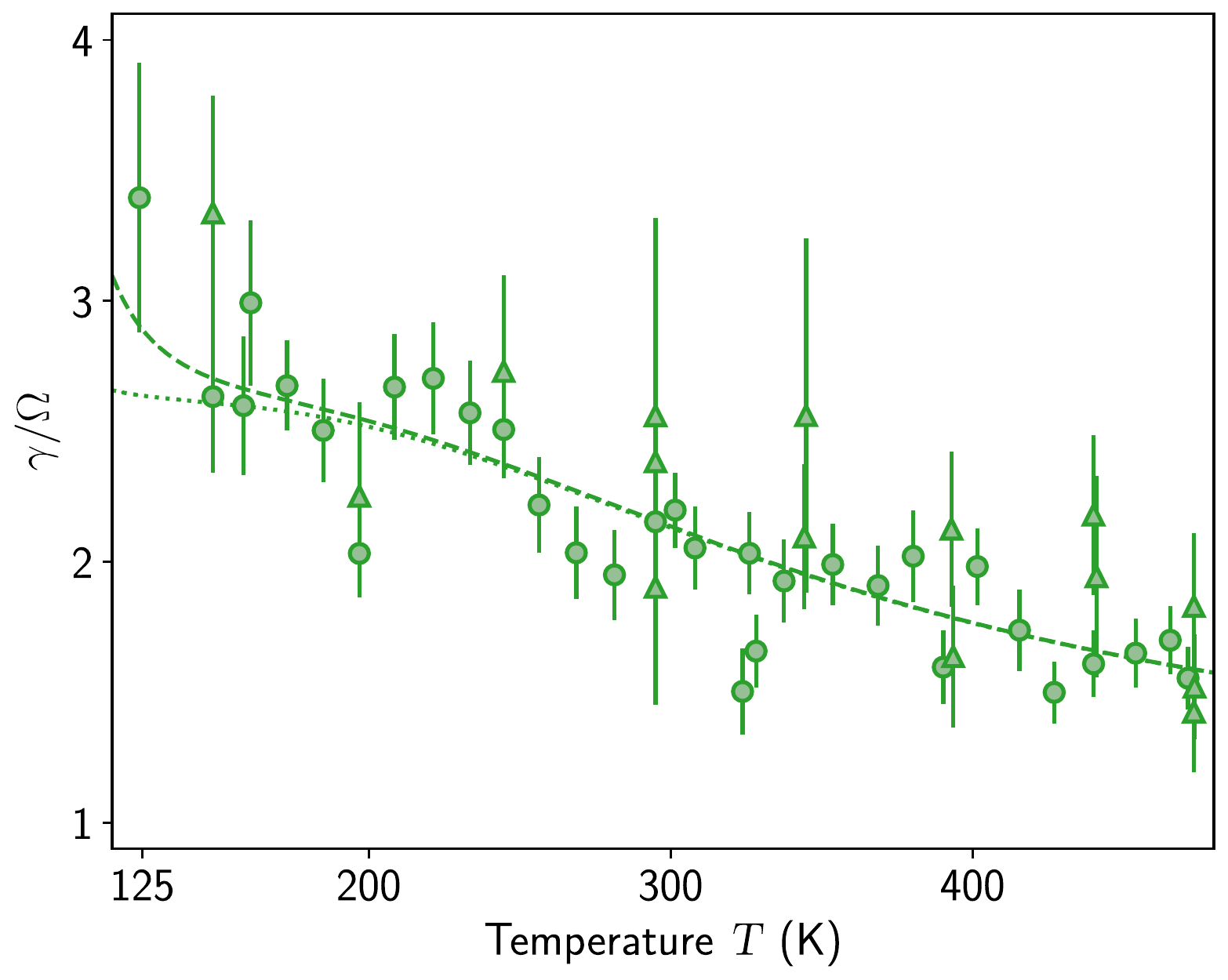}
\caption{\label{fig:rates_ratio}
Ratio of \(\gamma\) to \(\Omega\) as a function of temperature in the phonon-limited regime. The lines are the ratios of the fits to the measured relaxation rates \(\gamma\) and \(\Omega\) according to the proposed model described by Eqs.~(4) and (5) of the main text. As in Fig.~2 of the main text, the dotted (dashed) line includes sample-dependent constants for sample A (B). For both samples, the ratio passes near 2 at room temperature, as observed in Ref.~\cite{cambria2021state}. 
}
\end{figure}

\section{Ratio of \(\gamma\) to \(\Omega\) as a function of temperature}

In Ref.~\cite{cambria2021state}, we observed that \(\gamma \approx 2 \Omega\). We previously speculated that this factor of 2 may be fundamental, implying that the ratio should be independent of temperature in the phonon-limited regime. Fig.~\ref{fig:rates_ratio} demonstrates that this is not the case, as the ratio decreases nearly linearly from approximately 2.5 at 150 K to approximately 1.5 at the highest temperatures accessed in this work. The decreasing ratio indicates that the single-quantum transition is more strongly coupled to the higher-energy effective phonon mode than is the double-quantum transition.

\section{\textit{Ab initio} calculation of the spin-lattice relaxation rates}

We proceed with the \textit{ab initio} calculation of the spin-lattice relaxation rates by first computing the spin-phonon matrix elements up to second order, then applying Fermi's golden rule, and finally taking the continuum limit, converting sums over matrix elements into integrals over spectral functions. We note that we apply Fermi's golden rule with the random phase approximation such that interference terms are dropped.

\subsection{Calculation of the spin-phonon matrix elements}

In this section we describe our computational methods for evaluating the spin-phonon matrix elements for the NV center. While the present discussion is limited to the NV center for specificity and clarity, we emphasize that our approach can be easily generalized to other systems with zero field splittings. While the particular spin and symmetry of the NV center limit the number of matrix elements that must be calculated here, more complicated systems can also be treated using the same techniques at the expense of added computational complexity. Similarly, our discussion assumes that the zero field splitting arises from the spin-spin interaction, but analogous calculations may be carried out for systems with zero field splittings that results from the spin-orbit interaction.

We calculate the spin-phonon matrix elements using plane wave supercell density functional theory (DFT). We apply the VASP implementation~\cite{kresse93,kresse96} of the plane wave supercell DFT calculations with the projector augmented wave (PAW) method. The supercell consists of a 512-atom simple cubic diamond with a single negatively charged NV defect embedded in the center. We keep the optimized diamond lattice constant (3.567~\AA) fixed during the geometry optimization procedure of the defective supercell. The forces acting on the atoms are lower than $10^{-4}$~eV/\AA\ in the global energy minimum of the adiabatic potential energy surface (APES). We apply $\Gamma$-point sampling of the Brillouin-zone. We use a cutoff of 370~eV for the plane wave basis and a cutoff of 740~eV for charge augmentation in all calculations. The phonons together with the associated normal coordinates are calculated by building up the Hessian matrix as the first numerical derivative of the forces acting on the atoms, where the atoms are moved by $0.015$~\AA\ in each direction and the resultant APES is fit to a parabola around the global energy minimum. The zero field splitting tensor (or $D$-tensor) is calculated within the PAW-method~\cite{Bodrog_2013} as implemented by Martijn Marsman. In the global energy minimum of the APES, the $D$-constant is equal to $(3/2) D_{zz}$ after diagonalization of the $D$-tensor. Spin-polarized Perdew-Burke-Ernzerhof (PBE) functionals~\cite{func:PBE_96} are employed in these procedures.

The spin-phonon matrix elements are then obtained by exploiting the dependence of the $D$-tensor on the normal coordinates:
\begin{equation}
    \overleftrightarrow{D}(\boldsymbol{R})=\overleftrightarrow{D}(\boldsymbol{R}=0)+\sum_{i}\frac{\partial\overleftrightarrow{D}}{\partial R_{i}}\biggr|_{\boldsymbol{R}=0}R_{i}+\frac{1}{2}\sum_{ij}\frac{\partial\overleftrightarrow{D}}{\partial R_{i}\partial R_{j}}\biggr|_{\boldsymbol{R}=0}R_{i}R_{j} \text{,}
    \label{eq:D_Q}
\end{equation}
where a homebuilt script is used to extract the coefficients in Eq.~\eqref{eq:D_Q}. In order to evaluate the second-order derivatives, we consider only the diagonal terms which satisfy $i=j$ and distort the $C_{3v}$ symmetric atomic positions by all degenerate $e_x$, $e_y$ phonon modes of the supercell by $\sqrt{(\Delta\boldsymbol{R})^{2}}=0.1\;\text{\AA}\sqrt{\text{a.m.u.}}$. The second-order spin-flipping matrix elements $V_{+0}^{ll}$ and $V_{+-}^{ll}$ then determine the \(D\)-tensor according to the symmetry-adapted expression:
\begin{equation}
\begin{split}\overleftrightarrow{D}(\boldsymbol{R})= &\overset{{\textstyle \overleftrightarrow{D}(R=0)}}{\overbrace{\left(\begin{array}{ccc}
\!-\frac{1}{3}D\\
 & \!\!\!\!\!\!-\frac{1}{3}D\\
 &  & \!\!\!\!\!+\frac{2}{3}D
\end{array}\right)}}+\text{linear\;terms}+\left(\begin{array}{ccc}
\!-\frac{1}{3}\\
 & \!\!\!\!-\frac{1}{3}\\
 &  & \!\!\!+\frac{2}{3}
\end{array}\right)\sum_{i}3V_{00}^{ii}{R}_{i}^{2}+\\
+ & \sum_{l}V_{+-}^{ll}\left[\left(\begin{array}{ccc}
1\\
 & \!\!-1\\
 &  & \;\;\;
\end{array}\right)(X_{l}^{2}-Y_{l}^{2})+\left(\begin{array}{ccc}
 & 1\\
1\\
 &  & \;\;\;
\end{array}\right)2X_{l}Y_{l}\right]\\
+ & \sum_{l}\sqrt{2}V_{+0}^{ll}\left[\left(\begin{array}{ccc}
 &  & 1\\
 & \;\;\;\\
1
\end{array}\right)(X_{l}^{2}-Y_{l}^{2})+\left(\begin{array}{ccc}
\;\;\;\\
 &  & 1\\
 & 1
\end{array}\right)2X_{l}Y_{l}\right]+...
    \label{eq:D_symmetric}
\end{split}
\end{equation}
where $R_{i}$, $X_{l}$ and $Y_{l}$ are the dimensionless coordinates (\textit{not normal coordinates}) for the phonon mode at energy $\hbar \omega_i$ or $\hbar \omega_l$. We note that while the index $l$ only covers the $e$ modes once, the index $i$ covers all $a_1$, $a_2$, $e_x$, $e_y$ modes and thus runs over the $e$ modes twice. Therefore, we omit $e_x$'s degenerate partner $e_y$ modes from our calculations because their effect is the same. Eq.~\eqref{eq:D_symmetric} can be transformed into the spin-phonon interaction $\hat{V}$ by:
\begin{equation}
\begin{split}
\hat{V}=\overleftarrow{S}\overleftrightarrow{D}\overrightarrow{S}=D(\hat{S}_{z}^{2}-\frac{1}{3}S(S+1))+\text{linear\;terms}+\sum_{i}3V_{00}^{ii}\bigl(\hat{S}_{z}^{2}-\frac{1}{3}\hat{S}(\hat{S}+1)\bigr)\hat{R}_{i}^{2} \\
+\sum_{l}V_{+-}^{ll}\left[\bigl(\hat{S}_{x}^{2}-\hat{S}_{y}^{2}\bigr)(\hat{X}_{l}^{2}-\hat{Y}_{l}^{2})+\bigl(\hat{S}_{x}\hat{S}_{y}+\hat{S}_{y}\hat{S}_{x}\bigr)\bigl(\hat{X}_{l}\hat{Y}_{l}+\hat{Y}_{l}\hat{X}_{l}\bigr)\right]
\\
+\sum_{l}\sqrt{2}V_{+0}^{ll}\left[\bigl(\hat{S}_{x}\hat{S}_{z}+\hat{S}_{z}\hat{S}_{x}\bigr)(\hat{X}_{l}^{2}-\hat{Y}_{l}^{2})+\bigl(\hat{S}_{y}\hat{S}_{z}+\hat{S}_{z}\hat{S}_{y}\bigr)\bigl(\hat{X}_{l}\hat{Y}_{l}+\hat{Y}_{l}\hat{X}_{l}\bigr)\right]+...
\end{split}
    \label{eq:D_operators}
\end{equation}
where $\overleftarrow{S}=\bigl(\overrightarrow{S}\bigr)^{\dagger}=\bigl(\begin{array}{ccc} \hat{S}_{x} & \hat{S}_{y} & \hat{S}_{z}\end{array}\bigr)$. We expand the dimensionless coordinates in terms of the phonon creation and annihilation operators: $\hat{R}_{i}=(b_{i}^{\dagger}+b_{i})/\sqrt{2}$ and $\{\hat{X},\hat{Y}\}_{l}=(b_{\{X,Y\}l}^{\dagger}+b_{\{X,Y\}l})/\sqrt{2}$.

\begin{figure}[t]
\includegraphics[width=0.48\textwidth]{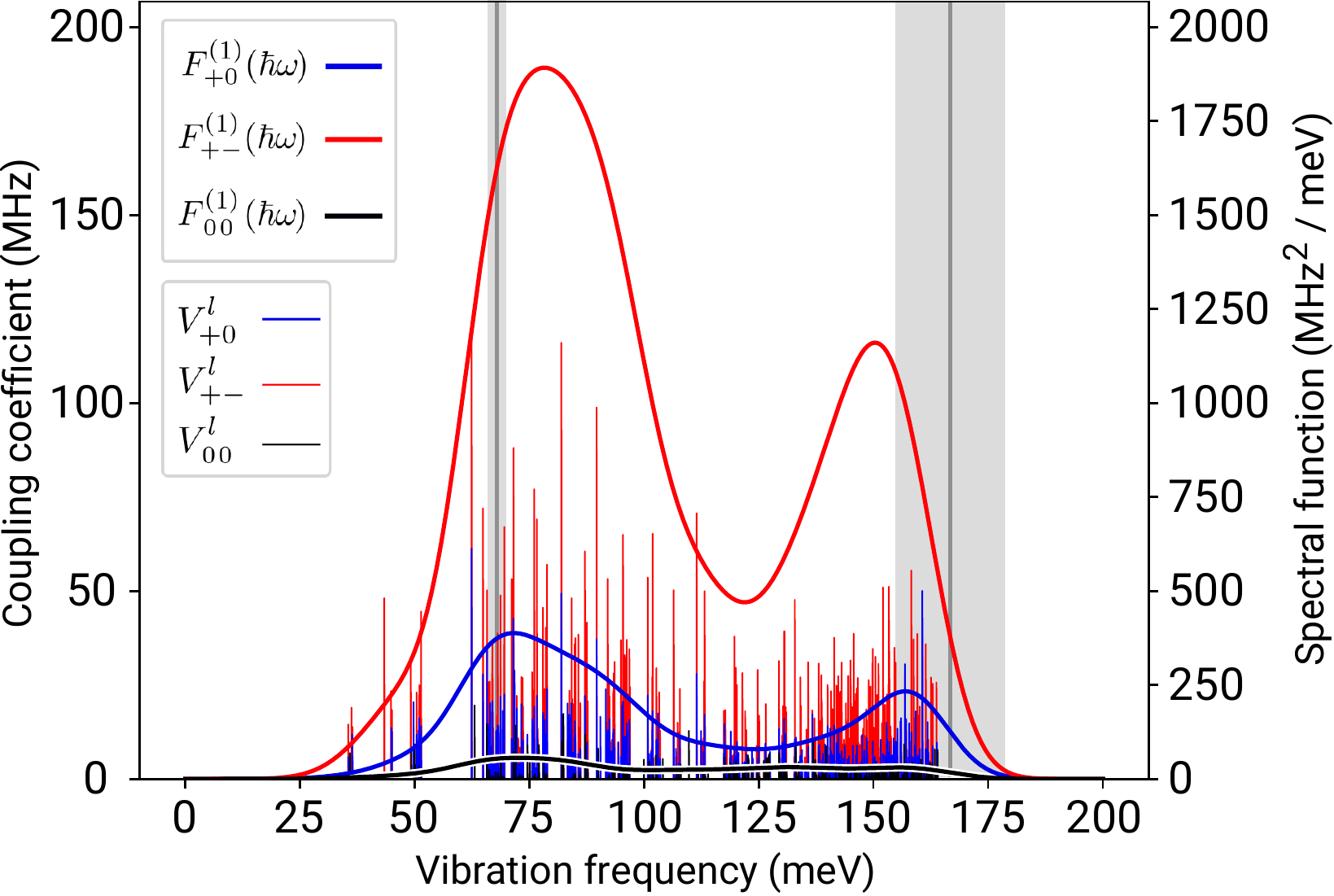}
\caption{\label{fig:first_spectral} 
\textit{Ab initio} calculation of the first-order spin-phonon coupling coefficients (thin lines) and spectral function (thick lines) for a single NV center in a 512 atom supercell. Like in Fig.~3 of the main text, we plot the coupling coefficients and spectral function for the magnitudes of the matrix elements \(\hat{S}_{z}\hat{S}_{+}\) (blue), \(\hat{S}_{+}^{2}\) (red), and \(\hat{S}_{z}^{2} - \tfrac{1}{3}\hat{S}^{2}\) (black). The first-order spectral function has the same double-peaked lineshape as the second-order spectral function. The spectral function is much smaller than the energies of the associated phonons, indicating that first-order interactions make negligible contributions to relaxation. As in Fig.~3 of the main text, the gray lines and confidence intervals mark the locations of the effective phonon modes as identified by the fit of the proposed model to the experimental data. 
}
\end{figure}

\subsection{Relaxation due to first-order spin-phonon interactions}

The first-order analog of Eq.~(2) in the main text describes the rate at which first-order interactions drive Raman transitions. According to Fermi's golden rule taken to second order in perturbation theory,
\begin{equation}
    \Gamma_{2(m_{s}m_{s}^{\prime})}^{(1)} (T) =
    \frac{2\pi}{\hbar}\sum_{ll^{\prime}m_{s}^{\prime\prime}}\left[
    n_{l}(n_{l^{\prime}}+1)\left|\frac{V_{m_{s}m_{s}^{\prime\prime}}^{l}V_{m_{s}^{\prime\prime}m_{s}^{\prime}}^{l^{\prime}}}{(E_{m_{s}''} - \hbar\omega_{l}) - E_{m_{s}}}\right|^{2}\delta(\Delta E_{-})
    +
    n_{l^{\prime}}(n_{l}+1)\left|\frac{V_{m_{s}m_{s}^{\prime\prime}}^{l}V_{m_{s}^{\prime\prime}m_{s}^{\prime}}^{l^{\prime}}}{(E_{m_{s}''} + \hbar\omega_{l}) - E_{m_{s}}}\right|^{2}\delta(\Delta E_{+})
    \right] 
    \text{,} \label{eq:gamma_12}
\end{equation}
where the \(V_{m_{s}m_{s}^{\prime}}^{l}\) are the first-order spin-phonon coupling coefficients from the first-order spin-phonon interaction $\hat{V}^{(1)}=\sum_{lm_{s}m_{s}^{\prime}}V_{m_{s}m_{s}^{\prime}}^{l}(a_{l}^{\dagger}+a_{l})$. As in the main text, $\Delta E_{\pm} = E_{m_{s}^{\prime}}\pm(\hbar\omega_{l}-\hbar\omega_{l^{\prime}}) - E_{m_{s}}$ is the energy difference between the final and initial states of the composite system. Making the approximation that the energy of the NV is small in comparison to the phonon energy,
\begin{equation}
    \Gamma_{2(m_{s}m_{s}^{\prime})}^{(1)} (T) =
    \frac{4\pi}{\hbar}\sum_{ll^{\prime}m_{s}^{\prime\prime}}
    n_{l}(n_{l^{\prime}}+1)\left|\frac{V_{m_{s}m_{s}^{\prime\prime}}^{l}V_{m_{s}^{\prime\prime}m_{s}^{\prime}}^{l^{\prime}}}{\hbar\omega_{l}}\right|^{2}\delta(\hbar\omega_{l'} - \hbar\omega_{l})
    \text{.} \label{eq:gamma_12_small_nv}
\end{equation}
In the continuum limit we replace the coupling coefficients with the first-order spectral function $F_{m_{s}m_{s}^{\prime}}^{(1)}(\hbar\omega)=\sum_{l}|V_{m_{s}m_{s}^{\prime}}^{l}|^{2}\delta(\hbar\omega-\hbar\omega_{l})$ and obtain
\begin{equation}
    \Gamma_{2(m_{s}m_{s}^{\prime})}^{(1)} (T) = \frac{4\pi}{\hbar}\sum_{m_{s}^{\prime\prime}} \intop_{0}^{\infty}d(\hbar\omega) n(\omega)[n(\omega)+1]\frac{F_{m_{s}m_{s}^{\prime\prime}}^{(1)}(\hbar\omega)F_{m_{s}^{\prime\prime},m_{s}^{\prime}}^{(1)}(\hbar\omega)}{(\hbar\omega)^{2}}. \label{eq:gamma_12_continous}
\end{equation}
We estimate the first-order spectral function for a macroscopic diamond by convolving the first-order coupling coefficients with a Gaussian of standard deviation \(\sigma=7.5 \text{ meV}\):
\begin{equation}
    F_{m_{s}m_{s}^{\prime}}^{(1)}(\hbar\omega) = \int d(\hbar\omega')\sum_{l} \abs{V_{m_{s}m_{s}^{\prime}}^{l}}^{2}\delta(\hbar\omega'-\hbar\omega_{l}) \times \frac{1}{\sigma \sqrt{2\pi}}\exp(-\frac{(\hbar\omega' - \hbar\omega)^{2}}{2\sigma^{2}})\label{first_conv}
\end{equation}
The \textit{ab initio} first-order coupling coefficients and spectral function are displayed in Fig.~\ref{fig:first_spectral}. Unlike for second-order interactions, Raman transitions involving first-order interactions are not directly related to the first-order matrix elements, but rather depend on products of the first-order matrix elements according to Eq.~\ref{eq:gamma_12_continous}. Nevertheless, we can estimate the first-order interaction contributions to relaxation by comparing the magnitude of the spectral function to the typical phonon energies. In validation of the argument presented in the main text, the \textit{ab initio} first-order matrix elements are on the order of MHz, much smaller than the energies of the associated phonons (\(\sim\)10 THz). We calculate that first-order interactions drive relaxation at rates several orders of magnitude slower than second-order interactions. 

\subsection{Relaxation due to second-order spin-phonon interactions}

From Fermi's golden rule taken to first order in perturbation theory, the second-order interaction $\hat{V}^{(2)}=\sum_{ll^{\prime}m_{s}m_{s}^{\prime}}V_{m_{s}m_{s}^{\prime}}^{ll^{\prime}}(a_{l}^{\dagger}+a_{l})(a_{l^{\prime}}^{\dagger}+a_{l^{\prime}})$ from Eq.~\eqref{eq:D_operators} drives Raman scattering at the rate shown in Eq.~(2) of the main text, which we reproduce here:
\begin{equation}
    \Gamma_{1(m_{s}m_{s}^{\prime})}^{(2)} (T) =
    \frac{2\pi}{\hbar}\sum_{ll^{\prime}}\left|V_{m_{s}m_{s}^{\prime}}^{ll^{\prime}}\right|^{2}\big[n_{l}(n_{l^{\prime}}+1)\delta(\Delta E_{-}) + n_{l^{\prime}}(n_{l}+1)\delta(\Delta E_{+})\big] \label{eq:secondspectral0}
\end{equation}
Again assuming that the NV energies are small in comparison to the phonon energies,
\begin{equation}
    \Gamma_{1(m_{s}m_{s}^{\prime})}^{(2)} (T) =
    \frac{4\pi}{\hbar}\sum_{ll^{\prime}}n_{l}(n_{l^{\prime}}+1)\left|V_{m_{s}m_{s}^{\prime}}^{ll^{\prime}}\right|^{2}\delta(\hbar\omega_{l'}-\hbar\omega_{l}) \label{eq:secondspectral1}
\end{equation}
We transform Eq.~\eqref{eq:secondspectral1} into its integral form:
\begin{equation}
\Gamma_{1(m_{s}m_{s}^{\prime})}^{(2)}(T)=\frac{4\pi}{\hbar}\intop_{0}^{\infty}\intop_{0}^{\infty}d(\hbar\omega)d(\hbar\omega^{\prime})n(\omega)(n(\omega)+1)F_{m_{s}m_{s}^{\prime}}^{(2)}(\hbar\omega,\hbar\omega^{\prime})\delta(\hbar\omega^{\prime}-\hbar\omega) \text{,} 
\label{eq:secondspectral2}
\end{equation}
by introducing the second-order spectral function $F_{m_{s}m_{s}^{\prime}}^{(2)}(\hbar\omega,\hbar\omega^{\prime})=\sum_{ll^{\prime}}|V_{m_{s}m_{s}^{\prime}}^{ll^{\prime}}|^{2}\delta(\hbar\omega-\hbar\omega_{l})\delta(\hbar\omega^{\prime}-\hbar\omega_{l^{\prime}})$. The $\delta(\hbar\omega^{\prime}-\hbar\omega)$ delta function in Eq.~\eqref{eq:secondspectral2} can be integrated out:
\begin{equation}
\Gamma_{1(m_{s}m_{s}^{\prime})}^{(2)}(T)=\frac{4\pi}{\hbar}\intop_{0}^{\infty}d(\hbar\omega)n(\omega)(n(\omega)+1)F_{m_{s}m_{s}^{\prime}}^{(2)}(\hbar\omega,\hbar\omega) \text{.}
\label{eq:secondspectral3}
\end{equation}
Following Eq.~\ref{first_conv}, we estimate the second-order spectral function for a macroscopic diamond by convolving the second-order coupling coefficients with a Gaussian of standard deviation \(\sigma=7.5 \text{ meV}\):
\begin{equation}
    \sqrt{F_{m_{s}m_{s}^{\prime}}^{(2)}(\hbar\omega,\hbar\omega)} = \int d(\hbar\omega')\sum_{l} \abs{V_{m_{s}m_{s}^{\prime}}^{ll}}\delta(\hbar\omega'-\hbar\omega_{l}) \times \frac{1}{\sigma \sqrt{2\pi}}\exp(-\frac{(\hbar\omega' - \hbar\omega)^{2}}{2\sigma^{2}}).\label{second_conv}
\end{equation}
We plot the square root of \textit{ab initio} spectral function in Fig.~3. of the main text. Ignoring degeneracies, we note that the $\hbar\omega^{\prime}=\hbar\omega$ constraint also enforces that $l=l^{\prime}$ and so we only consider the $V_{m_{s}m_{s}^{\prime}}^{ll}$ diagonal matrix elements in our \textit{ab initio} calculation.

\begin{figure}[t]
\includegraphics[width=0.98\textwidth]{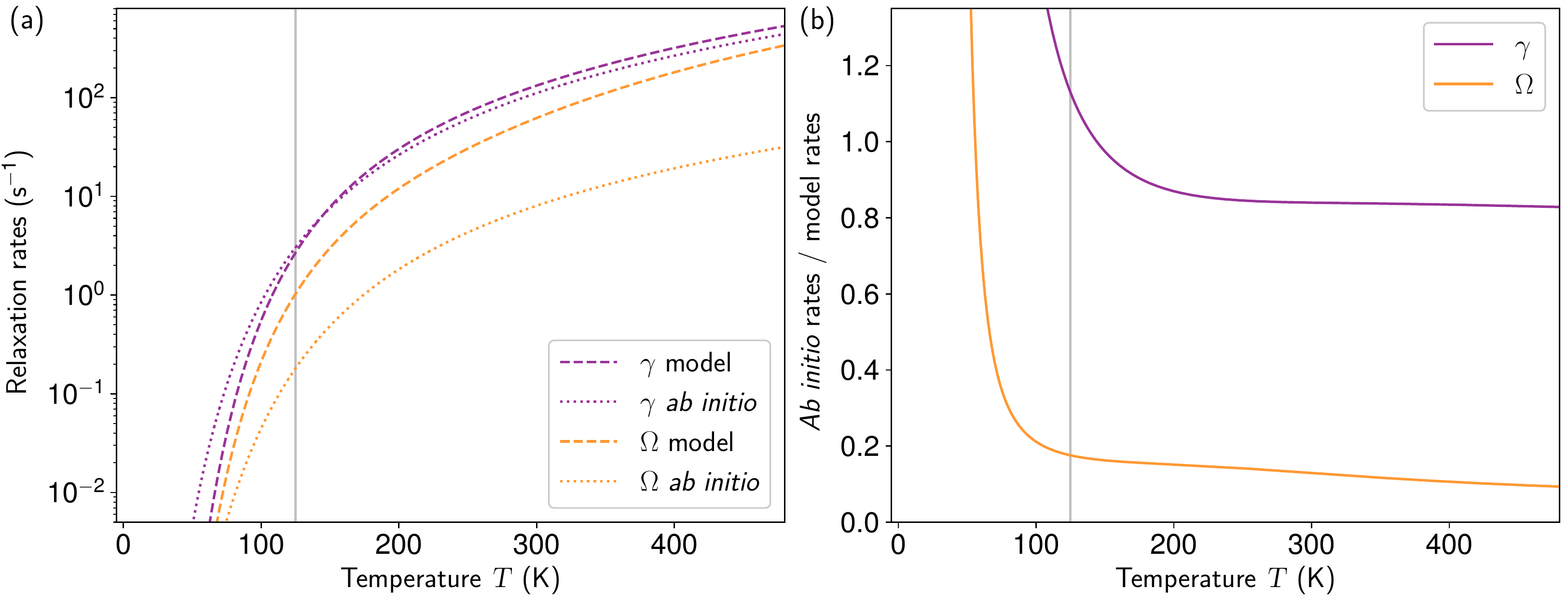}
\caption{\label{fig:ab_initio_rates}
Comparison between \textit{ab initio} theory and experiment. (a) Dotted lines show relaxation rates obtained by evaluating Eq.~(3) in the main text with the \textit{ab initio} second-order spin-phonon spectral function shown in Fig.~3 of the main text.  Dashed lines show fit of the analytical model (Eqs.~(4) and (5) in the main text) to the experimental data with sample-dependent constants set to 0. (b) Ratio of the \textit{ab initio} relaxation rates to the analytical model rates. In the phonon-limited regime (gray line) the \textit{ab initio} theory underestimates the experimentally measured relaxation rates by approximately 16\% for \(\gamma\) and a factor of 8 for \(\Omega\) at room temperature. 
}
\end{figure}

Fig.~\ref{fig:ab_initio_rates}a compares the relaxation rates predicted by the \textit{ab initio} theory in comparison to the fit of the analytical model to the experimental data. We obtain the predicted relaxation rates by evaluating Eq.~(3) of the main text using the calculated second-order spin-phonon spectral function for the single- and double-quantum transitions shown in Fig.~3 of the main text. 
The \textit{ab initio} theory achieves near quantitative agreement with experiment and captures several important features (Fig.~\ref{fig:ab_initio_rates}b). In particular, we observe that the theory reproduces the experimental finding that \(\gamma > \Omega\). The theory curves also closely follow the lineshapes of the experimental curves, as evidenced by the flat ratios between the \textit{ab initio} and model rates in the phonon-limited regime.

\subsection{Characteristics of the most strongly coupled phonon modes}

For a generic solid-state spin system that features a zero-field splitting, it is to be expected that the most strongly coupled phonon modes are those which induce large displacements of the specific atoms that primarily host the electronic wavefunction. We therefore predict that the most strongly coupled modes are also (quasi)localized modes for defects with highly localized electronic wavefunctions. Indeed, for the NV center we find that the most strongly coupled phonon modes are associated with large displacements of the three carbon atoms that support the unterminated ``vacancy lobes'' of the NV’s \(^{3}\text{A}_{2}\) electronic ground state triplet wavefunction. The most localized vibrational mode (~18\% probability on the three carbon atoms) with doubly degenerate ``E'' symmetry is at 62.4 meV and exhibits the largest spin-phonon coupling coefficients at 2 MHz for \(V_{+-}^{ll}\) and 0.6 MHz for \(V_{+0}^{ll}\), as can be seen in Fig. 3 of the main text. The most localized ``optical'' vibrational mode (11\% localization on the three carbon atoms) with doubly degenerate ``E'' symmetry is at 160.7 meV and exhibits spin-phonon coupling coefficients of 0.34 MHz for \(V_{+-}^{ll}\) and 0.07 MHz for \(V_{+0}^{ll}\). We note that it is not necessarily the case for other systems that the energies involved in the reduction of relaxation rate integral (Eq. 3 of the main text) to a limited sum (Eqs. 4 and 5 of the main text) will be associated with especially strongly coupled phonon modes. It may also be the case that the energies involved in the reduction are associated with a particularly large phonon density of states, and that there are no especially strongly coupled modes at those energies.

\begin{figure}[b]
\includegraphics[width=0.48\textwidth]{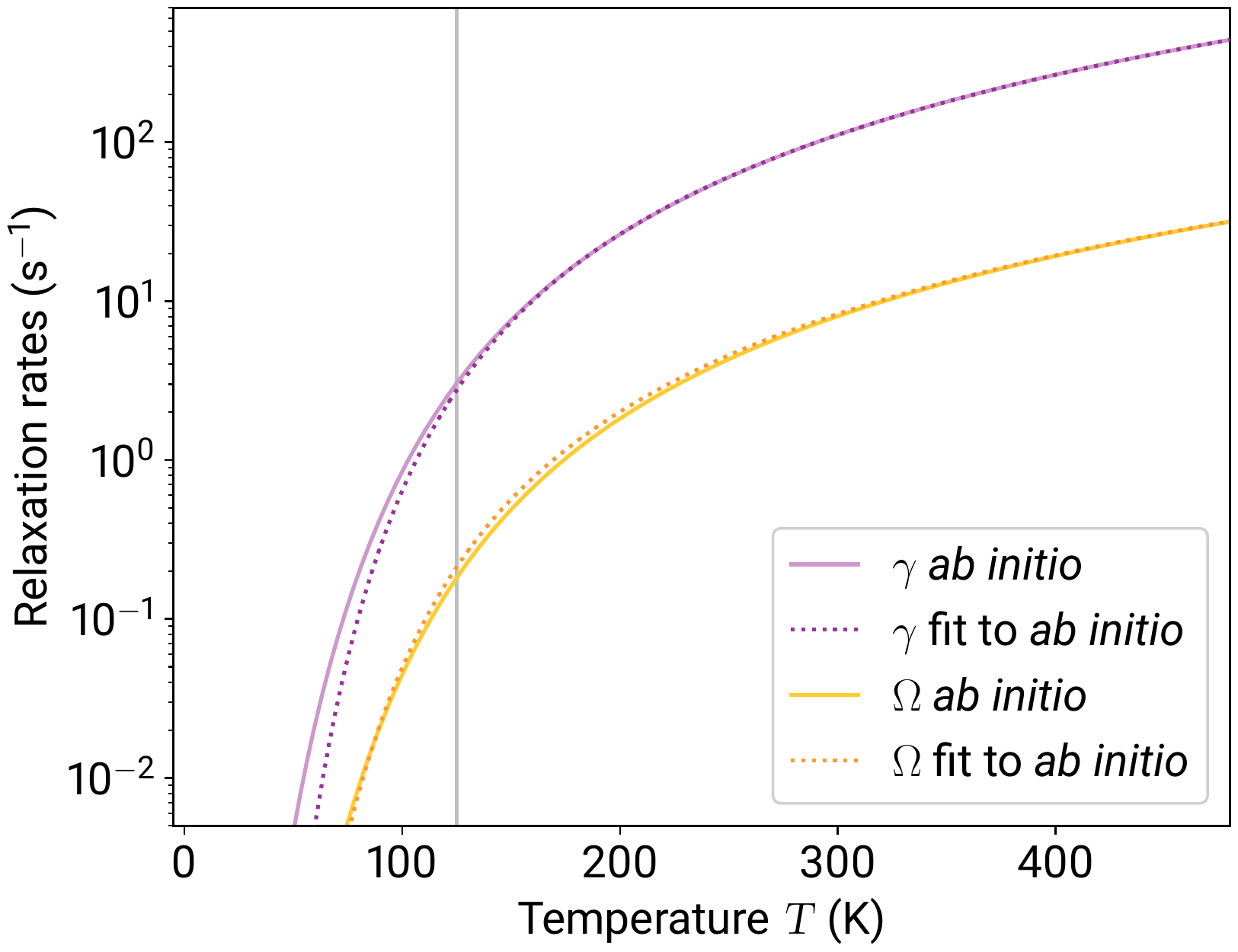}
\caption{\label{fig:ab_initio_fit}
Comparison of the relaxation rates predicted by \textit{ab initio} calculations (light, solid lines) and fits of the proposed model to the \textit{ab initio} rates (darker, dotted lines).
}
\end{figure}

\subsection{Fit of the proposed model to the \textit{ab initio} predicted relaxation rates}

In order to compare the salient characteristics of the experimental and \textit{ab initio} relaxation rates, it is useful to fit both data sets to the same model so that the fit parameters may be compared. Here, we fit a version of the proposed model without sample-dependent constants (i.e., a sum of two Orbach-like terms) to the relaxation rates predicted by the \textit{ab initio} simulations up to 5000 K. The results are shown in Fig.~\ref{fig:ab_initio_fit}, and the fit parameters are tabulated to 3 significant digits in Table~\ref{tab:ab_initio_fit}. The figure shows the ab initio predicted rates in lightly colored solid lines (matching the presentation from Fig. 2 of the main text) and the fit rates in darker dotted lines. Interestingly, we see that the activation energies extracted from this procedure are around 5-10\% lower than would be expected based on the spectral function and the fit of the proposed model to the experimental data. We attribute this effect to the larger occupation numbers, and thus larger contributions to the calculated relaxation rates, associated with lower-energy features of the spectral function. Using a toy model spectral function, we observe that the magnitude of the activation energy underestimation increases with the width of the Gaussian used for broadening the spectral function. We therefore anticipate that larger supercell calculations which allow for narrower Gaussian broadening may demonstrate less bias in the activation energies extracted from the fit.

\begin{table*}[t]
    \begin{tabularx}{0.8\textwidth}{|>{\centering\arraybackslash}X >{\centering\arraybackslash}X >{\centering\arraybackslash}X >{\centering\arraybackslash}X >{\centering\arraybackslash}X >{\centering\arraybackslash}X |} 
     \hline
     \(A_{1}\) (s\(^{-1}\)) & \(B_{1}\) (s\(^{-1}\)) & \(\Delta_{1}\) (meV) & \(A_{2}\) (s\(^{-1}\)) & \(B_{2}\) (s\(^{-1}\)) & \(\Delta_{2}\) (meV) \\
     \hline
     \num{0.0700e3} & \num{0.910e3} & 62.5 & \num{0.169e3} & \num{2.94e3} & 139 \\
     \hline
    \end{tabularx}
    \caption{Version of Table~\ref{tab:double_orbach} for the relaxation rates predicted by the \textit{ab initio} calculations. The \textit{ab initio} rates are fit according to the model described by Eqs. (4) and (5) of the main text without sample-dependent constants.}
    \label{tab:ab_initio_fit}
\end{table*}

\section{Generality of the theoretical arguments and methods}

The theory presented in this work can be generalized to other spin systems. In particular, second-order interactions will dominate Raman scattering in systems that have no low-lying excited states and feature phonons whose energies are much higher than the strengths of their respective couplings to the spin. We expect systems in which both the defect and lattice consist of light elements with strong covalent bonds to commonly meet these criteria. Other likely candidates are defects with ground state orbital singlets, due to the absence of the spin-orbit interaction. While our computational approach is not limited to spin-1 or \(C_{3v}\) symmetric systems, we highlight that our arguments and methods could be applied with little modification to several other spin-1 crystal defects with similar symmetries to the NV center, such as divacancy centers in silicon carbide~\cite{son2006divacancy, koehl2011nature} or the boron vacancy center in diamond~\cite{umeda2022negatively}. In contrast, the neutrally charged silicon vacancy center in diamond is likely not a good candidate due to the presence of a low-lying excited state \cite{rose2018strongly}, and the negatively charged silicon vacancy center in diamond and analogous defects involving elements from the same group are also likely poor candidates due to their large spin-orbit couplings \cite{hepp2014electronic}.

\section{Complete set of experimental data}

\begin{figure}[h]
\includegraphics[width=0.98\textwidth]{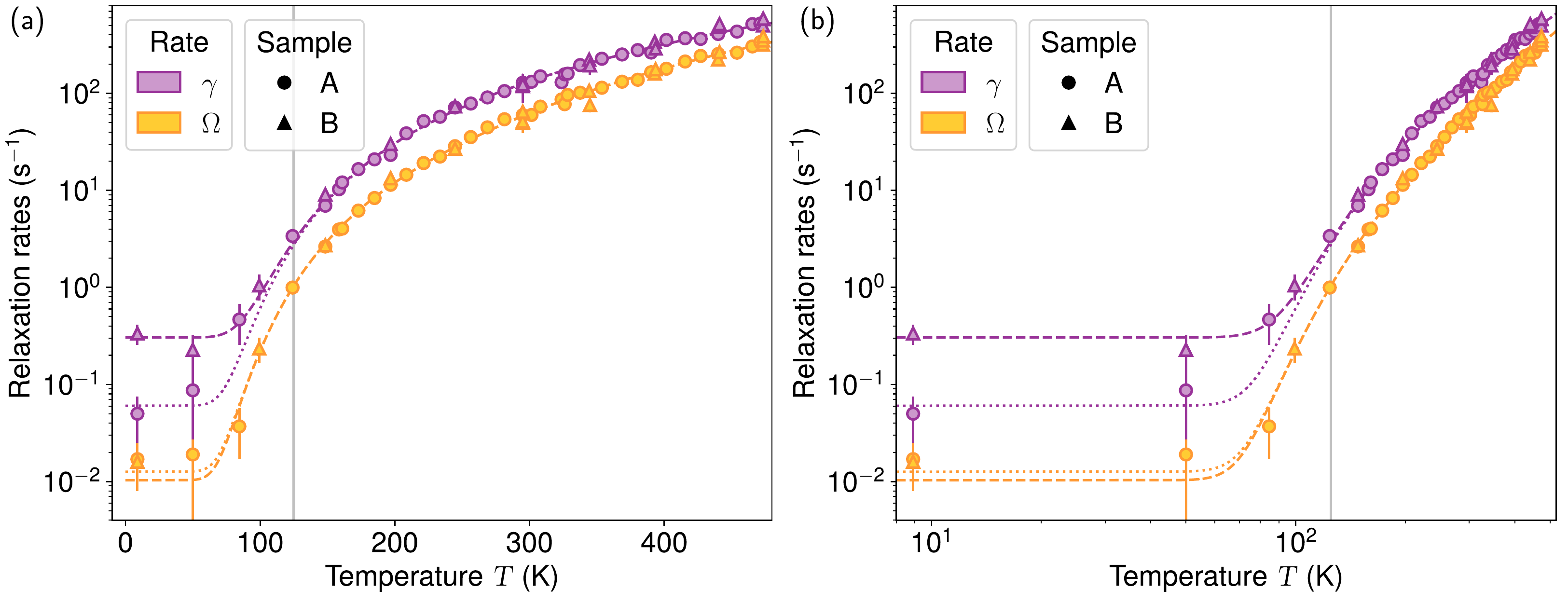}
\caption{\label{fig:rates_extended}
Semi-log and log-log versions (panels a and b respectively) of Fig.~2 from the main text showing the full temperature over which data was collected for this work. The rates predicted by the \textit{ab initio} calculation are not shown.
}
\end{figure}

Fig.~\ref{fig:rates_extended} and table~\ref{tab:complete_data} display the complete set of experimentally measured rates collected for this work. In the table, NVs are indexed first by sample and then numerically to distinguish single NVs in sample B.

\begin{table*}[b]
    \begin{tabularx}{\textwidth}{n t r r || n t r r} \\
     \hline
     \multicolumn{1}{c}{NV} & \multicolumn{1}{c}{\(T \text{ (K)}\)} & \multicolumn{1}{c}{\(\Omega\) (s\(^{-1}\))} & \multicolumn{1}{c||}{\(\gamma\) (s\(^{-1}\))} & \multicolumn{1}{c}{NV} & \multicolumn{1}{c}{\(T \text{ (K)}\)} & \multicolumn{1}{c}{\(\Omega\) (s\(^{-1}\))} & \multicolumn{1}{c}{\(\gamma\) (s\(^{-1}\))} \\
     \hline
        NVA & 8.9 & \num{1.7(9)e-2} & \num{5(3)e-2} & NVA & 390.3 & \num{1.65(7)e2} & \num{2.6(2)e2} \\
        NVA & 50.0 & \num{2(2)e-2} & \num{9(6)e-2} & NVA & 401.6 & \num{1.79(7)e2} & \num{3.5(2)e2} \\
        NVA & 84.6 & \num{4(2)e-2} & \num{5(2)e-1} & NVA & 415.6 & \num{2.13(10)e2} & \num{3.7(3)e2} \\
        NVA & 124.1 & \num{1.00(9)e0} & \num{3.4(4)e0} & NVA & 427.1 & \num{2.43(10)e2} & \num{3.6(2)e2} \\
        NVA & 148.4 & \num{2.64(16)e0} & \num{7.0(6)e0} & NVA & 440.1 & \num{2.54(11)e2} & \num{4.1(3)e2} \\
        NVA & 158.6 & \num{4.0(3)e0} & \num{1.03(8)e1} & NVA & 454.0 & \num{2.62(11)e2} & \num{4.3(3)e2} \\
        NVA & 160.9 & \num{4.0(3)e0} & \num{1.21(9)e1} & NVA & 465.5 & \num{3.04(13)e2} & \num{5.2(3)e2} \\
        NVA & 172.9 & \num{6.2(2)e0} & \num{1.66(8)e1} & NVA & 471.4 & \num{3.37(15)e2} & \num{5.2(3)e2} \\
        NVA & 184.9 & \num{8.3(4)e0} & \num{2.09(13)e1} & NVB3 & 8.9 & \num{1.6(6)e-2} & \num{3.4(8)e-1} \\
        NVA & 196.9 & \num{1.14(5)e1} & \num{2.32(16)e1} & NVB3 & 50.0 & \num{0.0(8)e-2} & \num{2.3(9)e-1} \\
        NVA & 208.5 & \num{1.45(7)e1} & \num{3.9(2)e1} & NVB2 & 99.3 & \num{2.3(7)e-1} & \num{1.1(3)e0} \\
        NVA & 221.4 & \num{1.91(10)e1} & \num{5.2(3)e1} & NVB2 & 148.4 & \num{2.7(2)e0} & \num{9.1(10)e0} \\
        NVA & 233.5 & \num{2.23(10)e1} & \num{5.7(4)e1} & NVB2 & 196.9 & \num{1.33(13)e1} & \num{3.0(4)e1} \\
        NVA & 244.7 & \num{2.85(13)e1} & \num{7.2(4)e1} & NVB2 & 244.7 & \num{2.7(2)e1} & \num{7.3(7)e1} \\
        NVA & 256.4 & \num{3.54(17)e1} & \num{7.9(5)e1} & NVB1 & 295.0 & \num{5.0(11)e1} & \num{1.2(4)e2} \\
        NVA & 268.8 & \num{4.5(2)e1} & \num{9.1(7)e1} & NVB5 & 295.0 & \num{5.1(5)e1} & \num{1.31(15)e2} \\
        NVA & 281.4 & \num{5.4(3)e1} & \num{1.05(8)e2} & NVB4 & 295.0 & \num{6.5(5)e1} & \num{1.23(13)e2} \\
        NVA & 295.0 & \num{6.0(3)e1} & \num{1.28(7)e2} & NVB5 & 344.3 & \num{1.06(9)e2} & \num{2.2(2)e2} \\
        NVA & 301.5 & \num{6.0(2)e1} & \num{1.32(7)e2} & NVB4 & 344.9 & \num{7.6(12)e1} & \num{1.9(4)e2} \\
        NVA & 308.1 & \num{7.3(3)e1} & \num{1.50(10)e2} & NVB5 & 393.0 & \num{1.60(13)e2} & \num{3.4(4)e2} \\
        NVA & 323.8 & \num{8.7(5)e1} & \num{1.30(12)e2} & NVB4 & 393.6 & \num{1.77(16)e2} & \num{2.9(4)e2} \\
        NVA & 326.1 & \num{7.7(3)e1} & \num{1.57(10)e2} & NVB5 & 440.1 & \num{2.23(19)e2} & \num{4.9(5)e2} \\
        NVA & 328.3 & \num{9.6(4)e1} & \num{1.59(11)e2} & NVB4 & 441.1 & \num{2.7(3)e2} & \num{5.2(8)e2} \\
        NVA & 337.6 & \num{1.01(5)e2} & \num{1.95(13)e2} & NVB5 & 473.3 & \num{3.2(3)e2} & \num{5.8(7)e2} \\
        NVA & 353.7 & \num{1.14(5)e2} & \num{2.27(15)e2} & NVB5 & 473.3 & \num{3.5(3)e2} & \num{5.0(7)e2} \\
        NVA & 368.7 & \num{1.32(6)e2} & \num{2.51(17)e2} & NVB4 & 473.5 & \num{3.9(3)e2} & \num{5.9(6)e2} \\
        NVA & 380.4 & \num{1.38(6)e2} & \num{2.8(2)e2} & & & \\
     \hline
    \end{tabularx}
    \caption{Complete set of experimentally measured rates collected for this work.}
    \label{tab:complete_data}
\end{table*}

\clearpage
\bibliography{SuppReferences.bib}